\documentclass[fleqn,usenatbib]{mnras}

\usepackage{afterpage}

\usepackage[T1]{fontenc}

\usepackage{graphicx}	
\usepackage{amsmath}	
\usepackage{amssymb}	
\usepackage{rotating}
\usepackage{pdflscape}
\usepackage[dvipsnames]{xcolor}
\usepackage{array, makecell}

\newcommand{\hb}{H$\beta$}
\newcommand{\lya}{Ly$\alpha$}

\newcommand{\hst}{\textit{HST}}
\newcommand{\msunyr}{M$_\odot$/yr}
\newcommand{\msun}{M$_\odot$}

\usepackage{newtxtext,newtxmath}

\title{Stellar feedback in a clumpy galaxy at $z \sim 3.4$}

\author[E. Iani et al.]{E. Iani$^{1,2,3}$\thanks{E-mail: iani@astro.rug.nl},
A. Zanella$^{4}$,
J. Vernet$^{3}$,
J. Richard$^{5}$,
M. Gronke$^{6}$\thanks{Hubble Fellow},
C. M. Harrison$^{7}$,
\newauthor
F. Arrigoni-Battaia$^{8}$,
G. Rodighiero$^{2}$,
A. Burkert$^{9,10}$,
M. Behrendt$^{9,10}$,
Chian-Chou Chen$^{11}$,
\newauthor
E. Emsellem$^{3,12}$,
J. Fensch$^{5}$,
P. Hibon$^{13}$,
M. Hilker$^{3}$,
E. Le Floc'h$^{14}$,
V. Mainieri$^{3}$,
\newauthor
A. M. Swinbank$^{15,16}$,
F. Valentino$^{17,18}$,
E. Vanzella$^{19}$
and M. A. Zwaan$^{3}$\\
Affiliations are listed at the end of the paper}

\date{Accepted 2021 July 27. Received 2021 July 8; in original form 2021 February 24}

\pubyear{2020}

\begin{document}
\label{firstpage}
\pagerange{\pageref{firstpage}--\pageref{lastpage}}
\maketitle

\begin{abstract}
Giant star-forming regions (clumps) are widespread features of galaxies at $z\approx1-4$.  
Theory predicts that they can play a crucial role in galaxy evolution if they survive to stellar feedback for $> 50$~Myr. Numerical simulations show that clumps' survival depends on the stellar feedback recipes that are adopted.
Up to date, observational constraints on both clumps' outflows strength and gas removal timescale are still uncertain. 
In this context, we study a line-emitting galaxy at redshift $z\simeq3.4$ lensed by the foreground galaxy cluster Abell 2895.
Four compact clumps with sizes $\lesssim 280$~pc and representative of the low-mass end of clumps' mass distribution (stellar masses $\lesssim 2\times10^8\ {\rm M}_\odot$) dominate the galaxy morphology. The clumps are likely forming stars in a starbursting mode and have a young stellar population ($\sim 10$~Myr). 
The properties of the Lyman-$\alpha$ (\lya) emission and nebular far-ultraviolet absorption lines indicate the presence of ejected material with global outflowing velocities of $\sim 200-300$~km/s.
Assuming that the detected outflows are the consequence of star formation feedback, we infer an average mass loading factor ($\eta$) for the clumps of $\sim 1.8 - 2.4$ consistent with results obtained from hydro-dynamical simulations of clumpy galaxies that assume relatively {\it strong} stellar feedback. Assuming no gas inflows (semi-closed box model), the estimates of $\eta$ suggest that the timescale over which the outflows expel the molecular gas reservoir ($\simeq 7\times 10^8\ \text{M}_\odot$) of the four detected low-mass clumps is $\lesssim 50$ Myr.
\end{abstract}

\begin{keywords}
galaxies: evolution -- galaxies: high-redshift -- galaxies: irregular -- galaxies: ISM -- galaxies: star formation 
\end{keywords}


\section{Introduction}
\label{sec:into}
Deep rest-frame ultraviolet (UV) and optical observations \cite[e.g.~][]{Driver+95, Glazebrook+95, vandenBergh+96, Driver+98, Elmegreen+07, Elmegreen+09, Overzier+10, Swinbank+10a, ForsterS+11a, Genzel+11, Wuyts+12, Guo+12, Conselice+14, Tadaki+14, Murata+14, Guo+15, Shibuya+16, Soto+17, Fisher+17a, Guo+18} have revealed that galaxies at the cosmic noon (redshift $z \sim 1-3$) typically display higher gas fractions \cite[][]{Daddi+10,Tacconi+10,Tacconi+13,Genzel+15}, star formation rates \cite[SFRs, e.g.][]{Genzel+06,ForsterS+06,Genzel+08}, and velocity dispersions \cite[][]{Elmegreen+05,ForsterS+06} than local star-forming galaxies.
Furthermore, bright concentrations of light, the so-called {\it clumps}, often dominate their light profile, thus making the clumps' host galaxies generally referred to as {\it clumpy galaxies}.

In the last decade, many efforts have been devoted to the understanding of clumps' nature and properties.
Clumps have been detected in rest-frame UV imaging \cite[e.g.~][]{Guo+12, Guo+15, Guo+18, Livermore+12, Shibuya+16, Soto+17, Dessauges-Zavadsky+18, Messa+19, Vanzella+21} as well as in maps of Balmer \cite[e.g. H$\alpha$, \hb, see~][]{Livermore+12, Mieda+16, Fisher+17a, Zanella+19, Whitmore+20} and Paschen \cite[e.g. Pa$\alpha$, see~][]{Larson+20} transitions. They also have been found to contribute to their host galaxies optical continuum \cite[e.g.~][]{Elmegreen+09, ForsterS+11a, ForsterS+11b} and CO emissions \cite[e.g.~][]{Jones+10, Dessauges-Zavadsky+17a}. 

Observations showed that clumps have sizes $\lesssim1$~kpc \cite[e.g.~][]{Elmegreen+07, ForsterS+11a}, estimated stellar masses (${\rm M}_\star$) of $\sim 10^7-10^9$~\msun\ \cite[e.g.~][]{ForsterS+11b, Guo+12, Soto+17}, and SFR from $0.1-10$~\msunyr\ \cite[e.g.~][]{Guo+12, Soto+17}.
Evidence also suggests that clumps are {\rm starbursting}, i.e. they have a specific star formation rate (sSFR$={\rm SFR/M}_\star$) that is a few orders of magnitude higher than the integrated sSFR of their host galaxies \cite[e.g.~][]{Bournaud+15, Zanella+15, Zanella+19}.
Because of these properties, clumps are therefore thought to trace giant star-forming regions. 


Several studies have highlighted how a comprehensive understanding of clumps could unveil the mechanisms driving star formation at high-redshift and provide critical insights on how galaxy assembly proceeds.
In particular, hydro-dynamical and cosmological simulations have suggested that if clumps survive to stellar feedback for hundreds of Myr \citep[e.g.][]{Gabor+13,Bournaud+11a,Bournaud+14,Mandelker+14,Mandelker+17}, while spiralling via dynamical friction towards the centre of the galaxy potential well, they generate torque and funnel inward large amounts of gas. With time, the inflow of gas contributes to the thickening of the galaxy disk and growth of the bulge \citep{Noguchi+99,Immeli+04a,Immeli+04b,ForsterS+06,Genzel+06,Genzel+08,Elmegreen+08,Carollo+07,Dekel+09a,Bournaud+09,Ceverino+10}, and possibly powers bright active galactic nucleus (AGN) episodes \citep{Bournaud+11b,Gabor+13,Dubois+12}. 
However, not all simulations agree with clumps survival scenario. Indeed, depending on the stellar feedback recipes adopted, clumps could retain much of their mass and survive \citep[{\it weak feedback}, e.g.][]{Immeli+04a,Elmegreen+08,Mandelker+14}, or be blown out by their own intense stellar feedback over timescales shorter than $\sim 50$ Myr \citep[{\it strong feedback}, e.g.][]{Murray+10,Genel+12,Hopkins+12,Tamburello+15,Buck+17,Oklopcic+17}. 
In this scenario, clumps' mass seems to play an important role since low-mass clumps are found to be affected by stellar feedback the most. 
It is therefore crucial to observationally constrain (as a function of clumps' stellar mass) the strength of stellar feedback (e.g. mass outflow rate, mass loading factor) as well as the timescale over which star formation consumes the gas reservoir and/or stellar winds and supernovae (SNe) outflows expel gas from the clumps.

\begin{figure}
    \centering
    \includegraphics[width=\columnwidth, trim=0cm .5cm 2.5cm 1cm, clip=true]{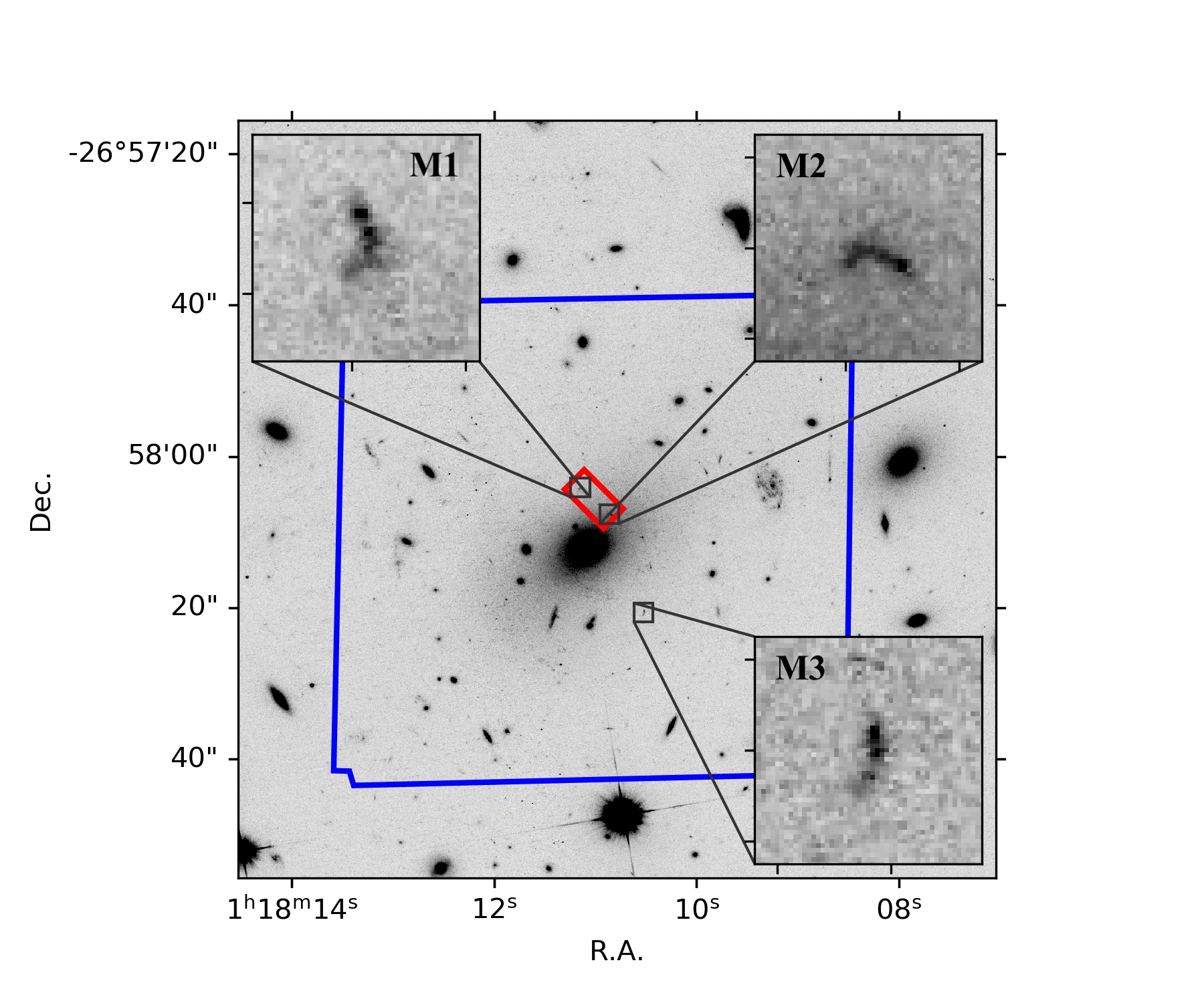}
    \caption{\textit{HST}/ACS WFC F606W image of the Abell 2895 galaxy cluster. We present cutouts for the three multiple images (M1, M2, M3) of our target A2895a \citep{Livermore+15}. Superimposed on the \hst\ image we display the contours of the 5h MUSE WFM+AO FoV (in blue), and the $\sim 5$h SINFONI K-band NoAO FoV (in red). The SINFONI observations cover only two of the three multiple images, i.e. M1 and M2.}
    \label{fig:multiple_images}
\end{figure}

In this framework, in this paper we investigate a high-redshift ($z\sim 3.4$) lensed (average magnification factor $\mu=7\pm1$) clumpy galaxy drawn from the sample of 12 gravitationally lensed galaxies by \cite{Livermore+15}.
We target a lensed galaxy since both lensing effects of magnification and stretching allow to reach very faint fluxes in a short amount of observing time and to spatially resolve galaxy substructures (e.g. clumps) down to sizes of $\sim 0.1$~kpc and, possibly, SFR~$\sim 1$~M$_\odot$/yr \cite[e.g.~][]{Jones+10, Livermore+12, Livermore+15, Rigby+17, Cava+18, Patricio+18, Dessauges-Zavadsky+18, Dessauges-Zavadsky+19}.
Our target (dubbed in \citealt{Livermore+15} as Abell 2895a) is lensed by the brightest cluster galaxy (BCG) residing at the very centre of the Abell 2895 (A2895, hereafter) galaxy cluster ($z\approx0.227$).
The galaxy has three multiple images (M1, M2, M3, see Figure~\ref{fig:multiple_images}) located at the celestial coordinates (right ascension, declination) of ($1^h 18^m 11.19^s$,$-26^\circ 58' 04.4''$), ($1^h 18^m 10.89^s$,$-26^\circ 58' 07.5''$), and ($1^h 18^m 10.57^s$,$-26^\circ 58' 20.5''$), respectively.
Thanks to the image multiplicity and lensing magnification, we are able to probe in detail the properties of this source from the multi-wavelength dataset at our disposal and composed of \hst, VLT/MUSE and SINFONI observations.

This paper is organised as follows.
In Section~\ref{sec:reduction}, we present our observations and data reduction.
In Section~\ref{sec:analyses}, we describe the lensing model of the A2895 galaxy cluster, discuss the morphological properties of our target, the method used to derive pseudo-narrow-band images of emission lines, extract the integrated far-ultraviolet (FUV) and optical spectra and the modelling of the target's \lya~emission.
From the FUV and optical spectra, in Section~\ref{sec:galaxy_prop}, we derive the galaxy physical properties (e.g. dust content, interstellar medium metallicity, SFR).  
Finally, in Section~\ref{sec:discussion}, we study clumps' gas outflows, and their properties. In particular, we derive the outflows energetic and clumps' gas removal timescale.
We summarise our results in Section~\ref{sec:conclusions}.

Throughout this paper, we adopt a Flat $\Lambda$-CDM cosmology with $\Omega_\Lambda = 0.7$, $\Omega_m = 0.3$, and $H_0 = 70\ \text{km s$^{-1}$ Mpc$^{-1}$}$.
When not differently stated, we assume a \cite{Chabrier+03} initial mass function (IMF)
and report all the measurements (e.g. lines flux, clumps' size) corrected for lensing effects.


\section{Observations and Data reduction}
\label{sec:reduction}
To study the rest-frame FUV and optical emission of our target galaxy, we gather a multi-wavelength dataset that combines archive \textit{HST} imaging and VLT/SINFONI near-IR integral-field spectroscopic data with new VLT/MUSE AO-assisted optical integral-field spectroscopy observations.
In the following, we describe the characteristics of each dataset and the procedure adopted for the data reduction.

\subsection{\textit{HST} data}
\label{subsec:hst}
The A2895 galaxy cluster was observed with the Advanced Camera for Surveys (ACS) on board \textit{HST} during Cycle 15 (SNAP program 10881, PI. G. Smith).
The observations were executed with the Wide Field Camera (WFC) F606W filter for a total exposure time of 0.33h.
The fully reduced F606W broad-band image was downloaded 
from the Hubble Legacy Archive\footnote{\url{https://hla.stsci.edu/}}.

To evaluate the point spread function (PSF) of the \textit{HST}/ACS image, we fit two dimensional (2D) gaussians to 5 non-saturated stars in the \textit{HST} field of view (FoV). The median value of the PSF full width at half maximum (FWHM) is $0.13''$.

To assess the absolute astrometry of the \textit{HST} image, we select 14 compact sources with high signal-to-noise ratio (SNR) and compare their \textit{HST} sky-coordinates with the GAIA DR2 catalogue \citep{Gaia+16,Gaia+18}.
We register the \textit{HST} astrometry to GAIA DR2 applying the inferred median-offsets of $\Delta \text{RA} = 0.66'' \pm 0.03''$ and $\Delta \text{Dec} = 0.06''\pm0.05''$.

\subsection{MUSE data}
\label{subsec:muse}
The central region of the A2895 galaxy cluster was observed with VLT/MUSE \citep{Bacon+10}, in Wide Field Mode with Ground-Layer Adapative Optics (GLAO) provided by the GALACSI module \citep{Arsenault+08,Strobele+12}.
The observations were carried out during the 2017 Science Verification of the MUSE AO module GALACSI (\citealt{leibundgut+17}; Programme ID: 60.A-9195(A), PI: A. Zanella), and in August 2019 (Programme ID: 0102.B-0741(A), PI: A. Zanella), for a total exposure time of 5 hours.
Each exposure was dithered and rotated by $90^\circ$ to obtain a combined dataset with more uniform noise properties.

We reduce the data via the ESO reduction (\textsc{esorex}) pipeline\footnote{\url{https://www.eso.org/sci/software/cpl/esorex.html}}, version 2.4.1 (\citealt{Weilbacher+20}).
We follow the standard reduction procedure, including bias correction, flat-fielding, wavelength and flux calibration, atmospheric extinction and astrometric correction. We disable the correction of telluric absorption since no suitable star in the field-of-view, nor a standard star close enough in time and airmass is available. Areas of strong telluric absorption are simply discarded in our analysis.
When combining individual exposures, we enable the use of the \textsc{autocalib} task that corrects for the background patterns caused by slightly different illumination of the MUSE slices. We remove sky residual lines from the final cube with the \textit{Zurich Atmosphere Purge} software (\textsc{zap}\footnote{\url{https://zap.readthedocs.io/en/latest/}} version 2.1, \citealt{Soto+16}).

As pointed out in \cite{Bacon+14}, the variance of the MUSE datacube propagated by the pipeline is underestimated.
To account for this, we define an extended region of the sky ($\sim 96~\text{arcsec}^2$) where we evaluate the sky flux variance at each wavelength.
We find that this is 1.35 times higher than the one estimated by the pipeline. We correct the MUSE variance cube by this factor. 

To bring the MUSE and \textit{HST} data on the same astrometric reference system, we consider 20 point-like sources selected from the MUSE white-light image.
Through bi-dimensional Gaussian modelling of the sources surface brightness profile, we evaluate their centroid celestial coordinates on both \textit{HST} and MUSE observations.
We correct the MUSE astrometry by adopting the median value of the difference between the GAIA-corrected \textit{HST} and MUSE celestial coordinates, i.e. $\Delta \text{RA} = -0.49'' \pm 0.07''$ and $\Delta \text{Dec} = -1.40'' \pm 0.08''$.

Finally, to reconstruct the MUSE PSF, we resort to \textsc{psfr}, the publicly available\footnote{\url{https://muse-psfr.readthedocs.io/en/latest/}} PSF reconstruction algorithm for MUSE data by \cite{Fusco+20}.
When applied to our observations, the \textsc{psfr} software retrieves a median PSF of $0.4''$ (FWHM).

\subsection{SINFONI data}
\label{subsec:sinfoni}
Our target galaxy was observed with the NIR integral-field spectrograph SINFONI \citep{Eisenhauer+03,Bonnet+04}, with the K-band grating, between July 25th and September 4th 2011 (Programme ID: 087.B-0875(A), PI: R. Livermore) without adaptive optics (NOAO mode). The total exposure time was 5.33h. 
The median seeing in the optical (at $\sim 6000$\AA) as evaluated by the telescope guide probe during the observations was $0.75''$ (FWHM).
Hence, the seeing-limited PSF of the observations in SINFONI K-band is $\sim 0.6''$ (FWHM).
 
We reduce the data via the ESO SINFONI pipeline (\textsc{esorex} version 3.13.2, \citealt{modigliani+07}) that corrects for dark current, bad pixels and distortions.
It also applies a flat field and performs a wavelength calibration.
We correct the science cubes for telluric features and flux calibrate them using the standard star observed before or after each observing blocks (OB).
The header astrometric information was used to combine science exposures within the same OB.
After the reduction of the single OBs, we correct their wavelength calibration for the barycentric velocity, a step that is not automatically performed by the pipeline. 
As the OBs were taken during different nights, we need to tie them to a common astrometric reference system before combining them in a final cube. To this aim, for each OB, we create an [OIII]$\lambda5008$ narrow-band image of the target, fit the emission with a 2D Gaussian, and estimate its centroid. 
We consider the [OIII]$\lambda5008$ emission of the target, as this is the brightest line at these wavelengths and the K-band continuum of the galaxy is not detected. 
Furthermore, the fact that the target shows two mirrored images (due to lensing effects) in the SINFONI field of view, helps us to accurately align the individual exposures.
We then mean-combine the cubes after applying a 3$\sigma$ clipping procedure to reject all spaxels affected by cosmic rays or displaying strong sky residuals.
Finally, we match the astrometry of the final cube with the \textit{HST} celestial coordinates, by minimising the spatial offset between the centroid of the [OIII]$\lambda5008$ emission and that of the \textit{HST} FUV continuum.
A geometrical reasoning supports this assumption: the distance between the two multiple images of [OIII]$\lambda5008$ matches the distance between the centroids of their FUV light.
Because of the mirroring effect of lensing, no offset along the direction orthogonal to the lensing critical line can be assumed.

\begin{figure*}
    \centering
    \includegraphics[width=\textwidth, trim = 0cm 1.4cm 0cm 0cm, clip = true]{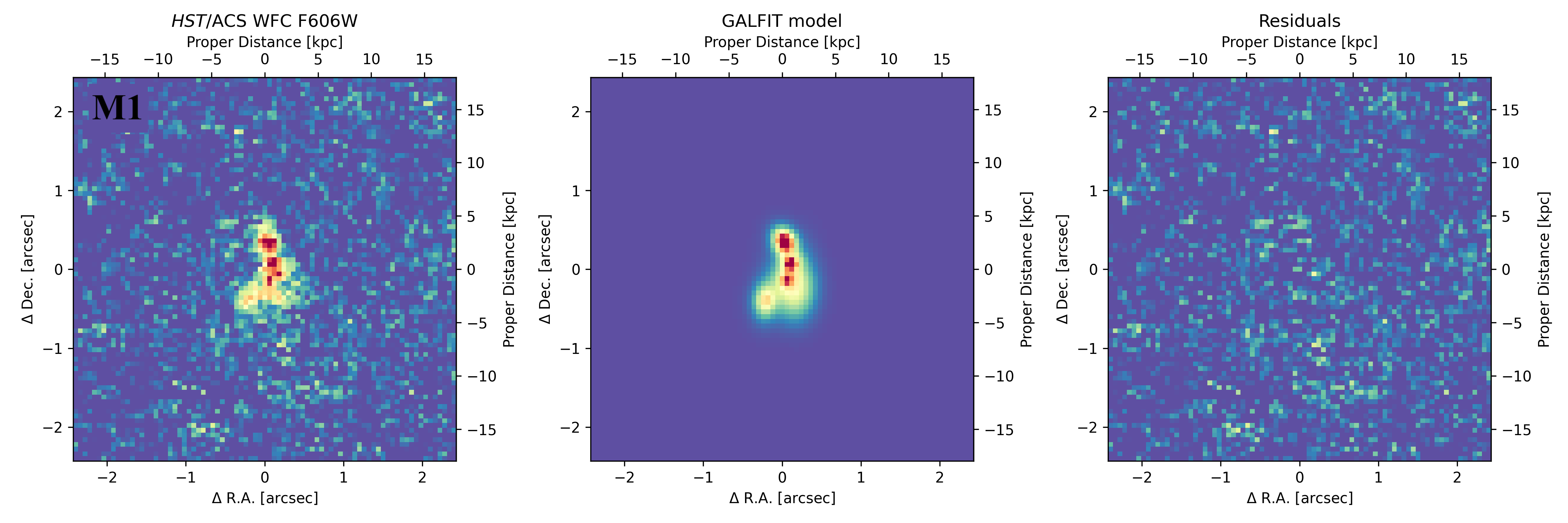}\\
    \includegraphics[width=\textwidth, trim = 0cm 1.4cm 0cm 1.8cm, clip = true]{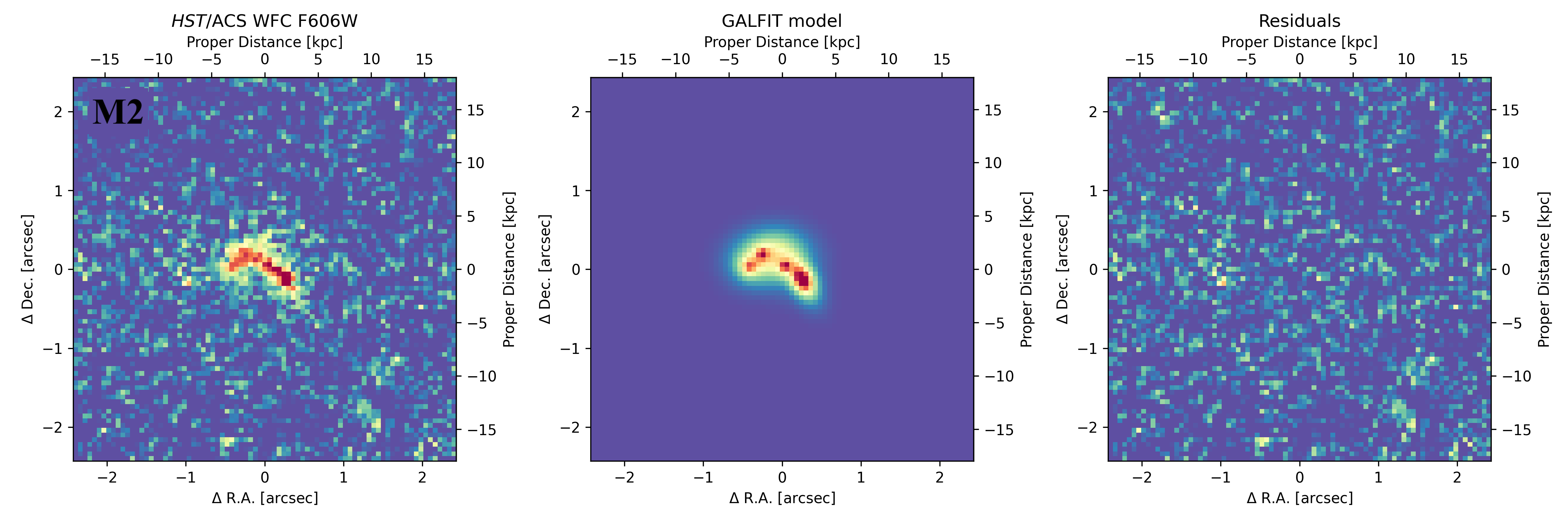}\\
    \includegraphics[width=\textwidth, trim = 0cm 0cm 0cm 1.8cm, clip = true]{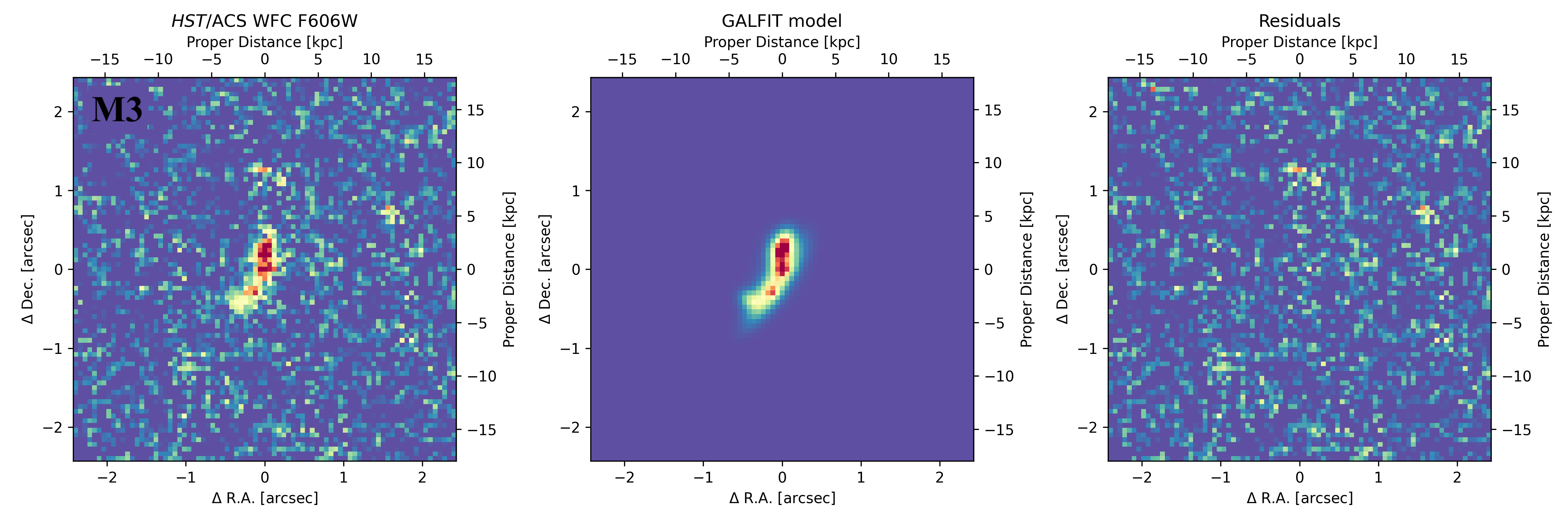}
    \caption{Results from the \textsc{galfit} 2D modelling of the three multiple images of our target, i.e. M1 (top), M2 (central) and M3 (bottom), on the galaxy image plane. The left panels show the galaxy light profile, as observed by \hst, after the subtraction of the A2895 BCG. The central panels display the best-fit \textsc{galfit} model (a diffuse component $+$ 4 clumps) while the panels on the right show the map of the residuals.}
    \label{fig:galfit_models}
\end{figure*}

\begin{figure*}
    \centering
    \includegraphics[width=.85\textwidth, trim = 1cm 0.25cm 3cm 1cm, clip = true]{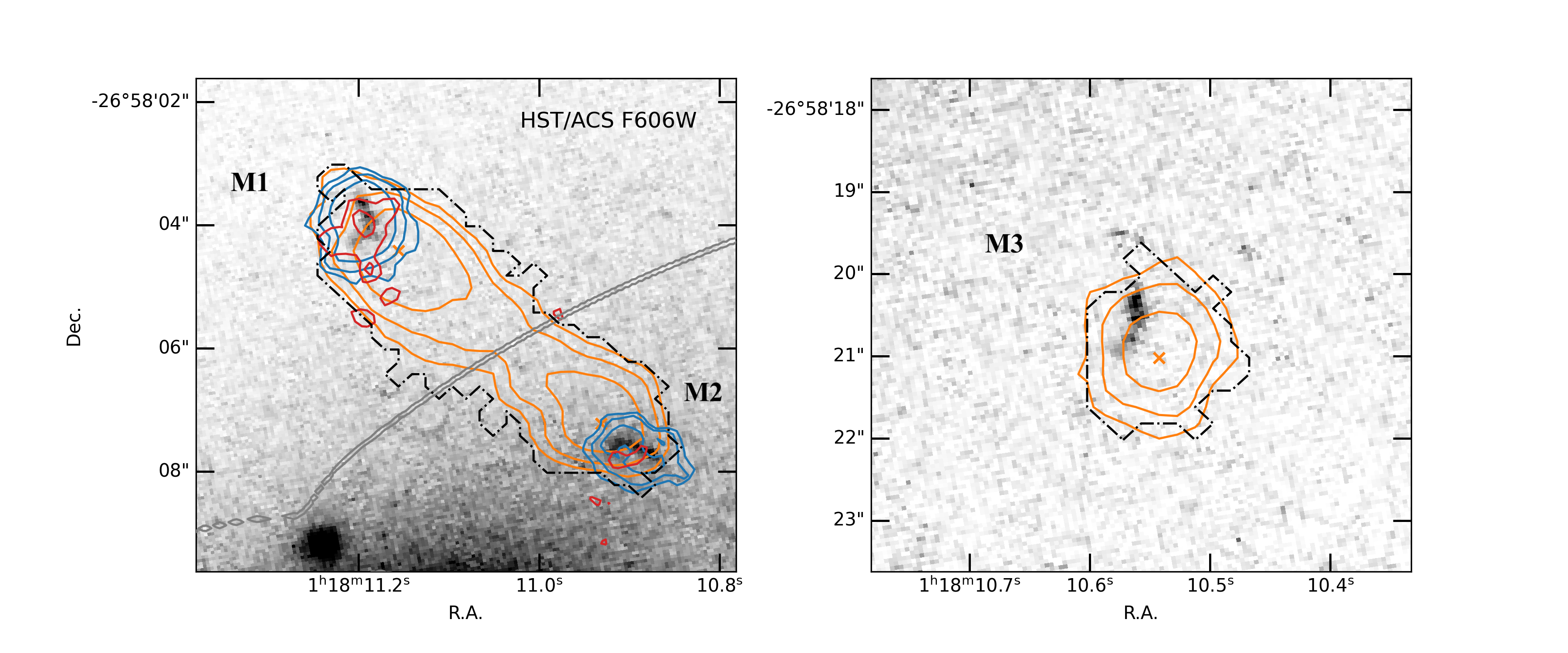}
    \caption{Spatial contours for the \lya~(orange), \hb~(red) and [OIII]$\lambda5008$ (blue) emissions for the three multiple images of our target galaxy, superimposed on the \textit{HST} imaging of the galaxy FUV continuum by \hst. The \lya\ and [OIII]$\lambda5008$ contour levels are at 2, 3 and 5$\sigma$ whereas, for \hb, they are at 1 and 2$\sigma$. The orange cross indicates the light-weighted position of the centroid of the \lya~emission. The grey solid line displays the lensing critical line. The absence of \hb~and [OIII]$\lambda5008$ contours in the right panel is because the M3 multiple image is not covered by SINFONI observations. Finally, the black dashed-dotted line shows the area within which we have extracted the target UV spectrum.}
    \label{fig:contours_emission_lines}
\end{figure*}

\begin{figure*}
    \centering
    \includegraphics[width=\textwidth]{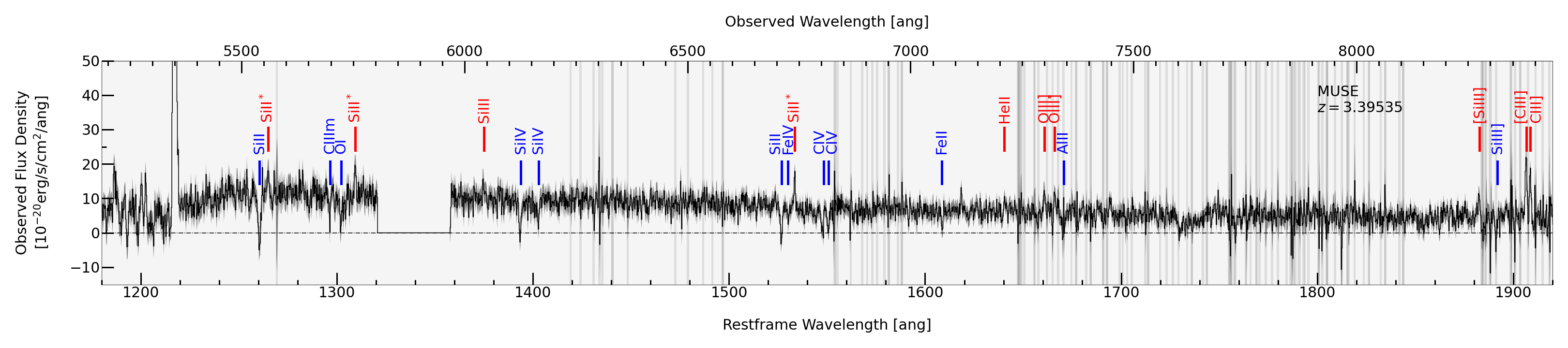}
    \includegraphics[width=\textwidth]{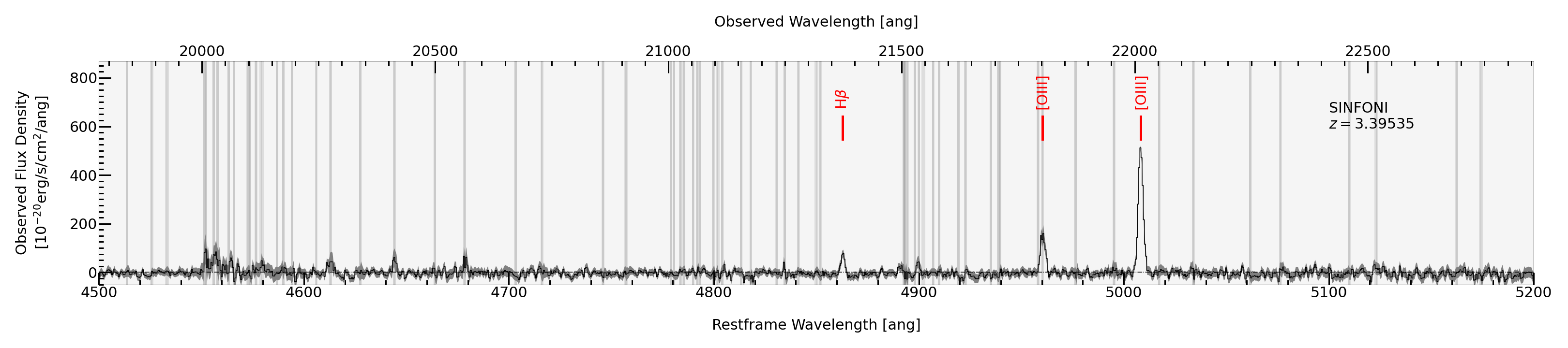}
    \caption{The rest-frame UV and optical spectra of our target galaxy. The grey-shaded regions display the $\pm1\sigma$ error around the spectra, while the vertical light-blue solid lines show the wavelength position of strong telluric lines.}
    \label{fig:spectra}
\end{figure*}


\begin{table*}
    \centering
    \begin{tabular}{lcrrrrrrr}
\hline
\hline
\thead{Line} & \thead{$\lambda_0^a$} & \thead{Flux$^b$} & \thead{SNR$^c$} & \thead{EW$_0^d$} & \thead{$z^e$} & \thead{$\sigma_\textnormal{obs}^f$} & \thead{$\sigma_\textnormal{corr}^g$}  & \thead{$\sigma^h$} \\
 & \thead{[\AA]} & \thead{[$10^{-20}$ erg/s/cm$^2$]} &   & \thead{[\AA]} &   & \thead{[\AA]} & \thead{[\AA]} & \thead{[km/s]} \\
\hline
SiII        & 1260.422     & -103.33  $\pm$ 8.95       & 14.1 & 2.1    $\pm$ 0.2    & -   & 2.6   $\pm$ 0.4     & 2.4  $\pm$ 0.4     & 127  $\pm$ 24    \\
SiII*       & 1264.738     & 28.70    $\pm$ 7.65       & 3.8  & -0.6   $\pm$ 0.2    & 3.39562 $\pm$ 0.00045   & 1.6   $\pm$ 0.5     & 1.2  $\pm$ 0.8     & 62   $\pm$ 41     \\
CIIIm       & 1296.330     & -24.94   $\pm$ 5.63       & 4.6  & 0.5    $\pm$ 0.1    & -   & 1.0   $\pm$ 0.3     & 0.5  $\pm$ 0.5     & 26   $\pm$ 27    \\
OI          & 1302.168     & -67.18   $\pm$ 11.14      & 6.3  & 1.4    $\pm$ 0.2    & -   & 4.0   $\pm$ 1.4     & 3.8  $\pm$ 1.5     & 201   $\pm$ 79    \\
SiII*       & 1309.276     & 37.69    $\pm$ 6.98       & 5.6  & -0.8   $\pm$ 0.1    & 3.39530 $\pm$ 0.00030   & 1.5   $\pm$ 0.4     & 1.0  $\pm$ 0.6     & 50   $\pm$ 28     \\
SiIII       & 1375.028     & 29.71    $\pm$ 6.78       & 4.5  & -0.7   $\pm$ 0.2    & 3.39464 $\pm$ 0.00061   & 2.7   $\pm$ 1.4     & 2.5  $\pm$ 1.6     & 124  $\pm$ 77     \\
SiIV        & 1393.755     & -58.48   $\pm$ 7.57       & 8.4  & 1.4    $\pm$ 0.2    & -   & 2.1   $\pm$ 0.6     & 1.8  $\pm$ 0.7     & 88   $\pm$ 34    \\
SiIV        & 1402.770     & -28.76   $\pm$ 6.70       & 4.4  & 0.7    $\pm$ 0.2    & -   & 1.5   $\pm$ 1.2     & 1.0  $\pm$ 1.7     & 50   $\pm$ 84    \\
SiII        & 1526.707     & -49.00   $\pm$ 4.14       & 14.7 & 1.5    $\pm$ 0.1    & -   & 1.9   $\pm$ 0.3     & 1.6  $\pm$ 0.3     & 71   $\pm$ 14    \\
FeIV        & 1530.040     & -13.38   $\pm$ 3.49       & 3.9  & 0.4    $\pm$ 0.1    & -   & 1.2   $\pm$ 0.8     & 0.5  $\pm$ 1.9     & 22   $\pm$ 86    \\
SiII*       & 1533.431     & 34.90    $\pm$ 5.48       & 6.7  & -1.1   $\pm$ 0.2    & 3.39542 $\pm$ 0.00025   & 1.3   $\pm$ 0.3     & 0.7  $\pm$ 0.6     & 29   $\pm$ 26     \\
CIV         & 1548.195     & -40.71   $\pm$ 6.48       & 6.6  & 1.3    $\pm$ 0.2    & -   & 2.3   $\pm$ 0.9     & 2.0  $\pm$ 1.0     & 90   $\pm$ 45    \\
CIV         & 1550.772     & -35.08   $\pm$ 5.99       & 6.1  & 1.2    $\pm$ 0.2    & -   & 2.2   $\pm$ 0.7     & 1.9  $\pm$ 0.8     & 83   $\pm$ 35    \\
FeII        & 1608.451     & -24.11   $\pm$ 3.11       & 8.4  & 0.8    $\pm$ 0.1    & -   & 1.5   $\pm$ 0.3     & 1.0  $\pm$ 0.5     & 43   $\pm$ 19    \\
HeII        & 1640.417     & 11.84    $\pm$ 3.71       & 3.2  & -0.4   $\pm$ 0.1    & 3.39556 $\pm$ 0.00038   & 1.1   $\pm$ 0.5     & 0.4  $\pm$ 1.4     & 18   $\pm$ 56     \\
OIII{]}     & 1660.809     & 24.86    $\pm$ 4.53       & 5.7  & -0.9   $\pm$ 0.2    & 3.39535 $\pm$ 0.00025   & 1.6   $\pm$ 0.5     & 1.2  $\pm$ 0.6     & 50   $\pm$ 25     \\
OIII{]}     & 1666.150     & 21.58    $\pm$ 4.75       & 4.7  & -0.8   $\pm$ 0.2    & 3.39531 $\pm$ 0.00026   & 1.4   $\pm$ 0.4     & 0.9  $\pm$ 0.5     & 39   $\pm$ 22     \\
AlII        & 1670.787     & -45.00   $\pm$ 9.61       & 4.8  & 1.7    $\pm$ 0.3    & -   & 3.9   $\pm$ 1.5     & 3.8  $\pm$ 1.5     & 153   $\pm$ 63    \\
{[}SiIII{]} & 1882.707     & 26.26    $\pm$ 4.54       & 6.1  & -1.3   $\pm$ 0.2    & 3.39512 $\pm$ 0.00025   & 1.5   $\pm$ 0.4     & 1.1  $\pm$ 0.6     & 38   $\pm$ 22     \\
SiIII{]}    & 1892.029     & -34.49   $\pm$ 5.99       & 6.0  & 1.7    $\pm$ 0.3    & -   & 2.4   $\pm$ 0.7     & 2.2  $\pm$ 0.8     & 80   $\pm$ 27    \\
{[}CIII{]}  & 1906.680     & 82.09    $\pm$ 6.73       & 15.4 & -4.1   $\pm$ 0.3    & 3.39515 $\pm$ 0.00025   & 1.8   $\pm$ 0.2     & 1.4  $\pm$ 0.3     & 51   $\pm$ 10     \\
CIII{]}     & 1908.734     & 53.97    $\pm$ 5.73       & 10.7 & -2.7   $\pm$ 0.3    & 3.39531 $\pm$ 0.00025   & 1.5   $\pm$ 0.3     & 1.1  $\pm$ 0.4     & 38   $\pm$ 13     \\
H$\beta$    & 4862.680     & 719.09   $\pm$ 161.50     & 9.8  & $\leq$ -3.5   $^\dagger$    & 3.39544 $\pm$ 0.00025   & 3.6   $\pm$ 0.7     & 3.3  $\pm$ 0.7     & 46   $\pm$ 10     \\
{[}OIII{]}  & 4960.295     & 2099.43  $\pm$ 437.22     & 17.2 & $\leq$ -9.4   $^\dagger$    & 3.39538 $\pm$ 0.00025   & 5.3   $\pm$ 0.5     & 2.1  $\pm$ 1.4     & 28   $\pm$ 19     \\
{[}OIII{]}  & 5008.240     & 6516.71  $\pm$ 1305.87    & 80.2 & $\leq$ -27.9 $^\dagger$    & 3.39527 $\pm$ 0.00025   & 5.2   $\pm$ 0.2     & 1.7  $\pm$ 0.5     & 23   $\pm$ 7     \\
\hline
    \end{tabular}
    \caption{Table of the results from the line fitting procedure presented in Section~\ref{subsec:lines_fit}. We report the parameters for the lines with a measured SNR $>3$. Unless differently stated, the measurements reported refer to the intrinsic values, i.e. corrected for lensing magnification. $^a$: the wavelengths reported are in vacuum. $^b$: the error on the flux has been increased by 5\% (MUSE) and 20\% (SINFONI) because of the error on the absolute calibration of the dataset. $^c$: the line SNR is estimated as the ratio between the flux and error measured from the fit only (i.e. without taking into account the additional absolute calibration error). $^d$: rest-frame EW of the line (the $^\dagger$ highlights lines for which the EW$_0$ has been estimated taking into account an upper limit on the stellar continuum flux). $^e$: estimated redshift of the target according to the wavelength of the best-fit Gaussian peak (only for nebular emission lines). $^f$: values of $\sigma$ obtained from the Gaussian fit. $^g$: values of the lines' $\sigma$ after the correction for instrumental broadening, i.e. $\sigma_{\rm corr} = \sqrt{\sigma_\textnormal{obs}^2-\sigma_\textnormal{instr}^2}$. For MUSE observations, we evaluate $\sigma_\text{instr}$ from the equation for the MUSE line spread function presented in \protect\cite{Bacon+17} (Equation 8).
    For SINFONI data, we adopt the value of 4.9\AA~(i.e. corresponding to two spectral pixels, see SINFONI user manual). $^h$: values of $\sigma_\text{corr}$ in units of km/s.}
    \label{tab:table_fitlines}
\end{table*}



\section{Analyses}
\label{sec:analyses}
In the following we report the procedures adopted to characterise the morphology of our target as well as its main properties.

\subsection{Lensing model}
\label{subsec:lensing_model}
The mass model we use in this work is constructed using the \textsc{lenstool}\footnote{\url{https://projects.lam.fr/projects/lenstool/wiki}} software \citep{Jullo+07}, following the methodology described in \cite{Richard+10}.
The 2D-projected mass distribution of the cluster is modelled as a parametric combination of one cluster-scale and several galaxy-scale double pseudo-isothermal elliptical potentials \citep{Eliasdottir+07}, representing the large-scale and cluster structure parts of the mass distributions, respectively.
To restrain the number of parameters in the model, the centres and shapes of the galaxy-scale components are constrained to the centroid, ellipticity and position angle of cluster members as measured on the \textit{HST} image.
The cluster members are assumed to follow the Faber-Jackson relation for elliptical galaxies \citep[][]{Faber+76}, and are selected through the colour-magnitude diagram method \citep[e.g.][]{Richard+14}.
This parametric model is constrained by using the location of two triply imaged systems with spectroscopic redshift and presented in \cite{Livermore+15}, i.e. A2895a and A2895b.
The best fit model reproduces the location of the multiply-imaged systems with an rms of $0.09''$. We use \textsc{lenstool} with this best fit parameters to produce a 2D map of the magnification factor at the redshift of A2895a.
We resample the maps of the lensing magnification to match the \hst, MUSE and SINFONI spatial sampling, respectively.
As a final step, we reconstruct the \hst~multiple images of A2895a on the galaxy source plane.
This is done using our lens model to raytrace back each spaxel observed in every multiple image, and subtract the lensing displacement.

\subsection{Galaxy morphology}
\label{subsec:morphology}
The FUV continuum probed by \textit{HST} shows that our target has an irregular morphology, dominated by bright star-forming regions, see cutouts from Figure~\ref{fig:multiple_images}.
The presence of substructures with an intrinsic effective radius ranging from 60 ($\sim 0.008''$) to 500 pc ($\sim 0.07''$) was already identified in \cite{Livermore+15} in the reconstructed SINFONI \hb~emission line map, despite the observed PSF (FWHM $\sim 0.6''$, corresponding to an intrinsic FWHM of $\sim 0.2''$ on the source plane).
To avoid possible bias induced by the use of reconstructed line maps on the galaxy source plane, we look for clumps directly on the image plane, leveraging the dataset with the highest angular resolution, i.e. \textit{HST} (observed PSF FWHM $\sim 0.13''$, $\sim0.04''$ in the source plane). 

To identify the clumps and understand what is their contribution to the overall galaxy emission, we implement an iterative modelling of the galaxy 2D surface brightness profile by means of the \textsc{galfit} software \citep{Peng+10}. 
The methodology we use follows the one presented in \cite{Zanella+19} but is tailored to our scientific case, i.e. it is applied to all the three multiple images of our target (M1, M2, M3) and requires the additional modelling of the A2895 BCG optical light gradient that contaminates the FUV emission of our galaxy.
We model the BCG 2D light profile by using two S\'ersic models. The first component fits the BCG extended disk (R$_\mathrm{e} \sim 100$ kpc, n $\sim 2$, consistent with the measurements reported by \citealt{Stott+11}); the second fits a central, more compact component (R$_\mathrm{e} \sim 5$ kpc, n $\sim 2$). After the subtraction of the BCG light profile, the background at the location of the multiple images of our target is well subtracted. 
We then model our target employing a 2D Gaussian profile. The map of the residuals highlights the presence of four clumps. Hence, we re-run \textsc{galfit} adding to the 2D Gaussian model of the overall galaxy (hereafter, the diffuse component) four additional 2D profiles each intended to represent a clump. 
The best fit of our target with minimum and non-structured residuals (see Figure~\ref{fig:galfit_models}) is obtained with a 2D Gaussian profile for the diffuse component, three 2D PSF and a 2D S\'ersic models for the clumps. 
Indeed, while three clumps out of four are unresolved and well reproduced by a PSF-like profile, one is marginally resolved, having a radius $\sim 0.10''$ (S\'ersic profile).
We repeat this analysis independently on the three images of our target and reach similar conclusions.

The \hst\ PSF gives us an upper-limit on the clumps' size of $\sim 280$ pc (value corrected for magnification) in radius. 
By summing the flux of all the clumps and comparing it with the total emission (clumps plus diffuse component), we conclude that $\sim$ 60\% of the FUV light is emitted by the four star-forming regions. 
To verify that our result is not biased by the choice of the \textsc{galfit} models used to fit the different components of the FUV emission (i.e. S\'ersic, PSF), we carry out an independent test based on the construction of the galaxy curve of growth, see Appendix~\ref{appendix:growth_curve}.
The results of this test confirm the \textsc{galfit} findings. 

We assume that, similarly to the FUV continuum, clumps also dominate the FUV and optical line emission. 
This is a reasonable assumption, given that the emission lines probed by the MUSE and SINFONI data trace star formation, similarly to the FUV continuum. 
Likely, the contribution of young clumps (age $\sim 10$~Myr , see Section~\ref{subsec:stellar_age_imf}) to the emission lines is even higher than the 60\% estimated for the continuum \citep{Zanella+19}.

\subsection{Emission lines pseudo-NB image}
\label{subsec:pseudo_nb}
As revealed by a first inspection of the MUSE and SINFONI observations, the FUV and optical spectra of our target feature several emission lines among which the brightest are \lya, \hb~and the [OIII]$\lambda4960,5008$ doublet ([OIII]db hereafter). 

To investigate the spatial extent of these emission lines, and to compare them with the FUV continuum from \hst, we create pseudo-NB images that maximise the lines' SNR, see Appendix~\ref{appendix:pseudo_NB_images}.
We extract the flux and variance spectra within circular apertures of increasing size (from $0.3-3.0''$, in steps of $0.2''$) centred at the position of each multiple image.
Then, we convolve each spectrum with Gaussians of increasing $\sigma$ (from $1.25-10$\AA, in steps of $1.25$\AA), and compute the SNR as a function of wavelength.
From the convolved spectrum that maximises the line SNR, we derive the peak position of the line ($\lambda_{\rm max}$) as well as its standard deviation ($\sigma_{\rm max}$).
The values obtained for the three multiple images are consistent with each other.
Hence, we define the wavelength range within which we collapse the datacube as given by the interval $\lambda_{\rm max} \pm 3\sigma_{\rm max}$.
However, before obtaining the pseudo-NB image, we subtract spaxel by spaxel any eventual continuum emission by fitting the spectral region adjacent to the line. 
Finally, we reconstruct the derived pseudo-NB image of each line on the galaxy source plane, following the same procedure as adopted for the \hst\ FUV continuum, see Section~\ref{subsec:lensing_model}.

In Figure~\ref{fig:contours_emission_lines}, we present the \lya, \hb~and [OIII]$\lambda5008$ emission contours overlaid on the rest-frame FUV \textit{HST} image.
While the H$\beta$ and [OIII] emission regions are spatially coincident with the FUV stellar continuum, the peak of \lya~is offset.  
To evaluate the displacement ($\delta_{{\rm Ly}\alpha}$) between the \lya~and the centroid of the galaxy FUV light, we model with 2D Gaussian profiles the emissions on the reconstructed map of the galaxy counter-image, i.e. the least stretched and magnified image of our target (M3), in the source plane. 
From the reconstructed map, we measure a \lya-UV intrinsic offset of $0.16''\pm0.02''$ that corresponds to $1.2\pm0.2$~kpc.
We resort to the reconstructed map of the galaxy counter-image since the \lya~haloes in the other two multiple images are incomplete and merged together. 
The \lya~emission appears to be extended and isotropic, i.e. without evidence of any clear substructure, at the resolution of our MUSE data. 
Despite the fact that offsets between the \lya\ and UV continuum of galaxies have been widely reported in the literature \cite[e.g.][and references therein]{Shibuya+14b, Hoag+19}, the origin of these displacements remains unclear. 
3D models of \lya\ radiative transfer \cite[e.g.][]{Laursen+07, Verhamme+12, Behrens+14, Zheng+14} of disk systems suggest that the \lya-UV offset could be ascribed to the easier propagation and escape of \lya\ photons in the direction perpendicular to the galaxy disk. 
Indeed, because of the resonant nature of \lya\ photons that makes them prone to undergo many scattering events, the distribution of neutral hydrogen and dust strongly affects the observed \lya\ distribution. 
In this case, the offset would be a consequence of the viewing angle under which the observer sees the target.
The offset estimate we find is in good agreement with the typical displacements reported in the literature for LAEs and Lyman-break galaxies (LBGs), i.e. $\delta_{{\rm Ly}\alpha}=1-4$~kpc \cite[e.g.][]{Bunker+00, Fynbo+01, Shibuya+14b, Hoag+19}.

\subsection{FUV and optical spectrum extraction}
\label{subsec:spectra_extraction}
To define the spatial regions of the MUSE and SINFONI datacubes where to extract the FUV and optical spectra of the galaxy, we resort to the \lya~and [OIII]$\lambda5008$ pseudo-NB images, i.e. the brightest lines of the FUV and optical dataset, respectively. 
For both line maps, we measure the background level and variance ($\sigma$)\footnote{For the MUSE data we consider the variance cube produced by the pipeline and corrected it as described in Section~\ref{subsec:muse}.
The SINFONI pipeline instead does not return a variance cube and therefore we evaluate, at each wavelength, the standard deviation of all the spaxels that do not show emission from the target.}, and define the area where to extract the galaxy spectrum as given by all the spaxels where the line flux is $\geq 2.5\sigma$.
The MUSE FUV spectrum however is heavily contaminated by the optical stellar continuum of the A2895 BCG.
To obtain a `clean' spectrum of our target we proceed as follows.
We mask all the sources around the A2895 central galaxy, including the spaxels belonging to our target.
For each MUSE spaxel with \lya~flux $>2.5\sigma$, we estimate its elliptical angular distance from the BCG centre, consider all the unmasked spaxels laying at the same distance, and create a median-combined spectrum of the BCG.
This spectrum is then subtracted from the original observed spectrum of our target.
In this way, we can effectively decontaminate it from the contribution of the BCG optical light.
We avoid to simply use a combined spectrum of the innermost regions of the BCG since we detect variations in the BCG spectrum as a function of its radius. 
After correcting each spaxel for its lensing magnification factor, we sum all the spectra corresponding to the spaxels with flux $\geq 2.5\sigma$.
Finally, we average the spectrum of all the available multiple images of our target to obtain a spectrum of maximum SNR. 
In Figure~\ref{fig:spectra} we present the FUV (upper panel) and optical (lower panel) spectra of our target.

\subsection{Emission and absorption line measurements}
\label{subsec:lines_fit}
Besides strong \lya, \hb~and [OIII]db, we detect a plethora of other FUV and optical lines (both in emission and absorption). 
To estimate their peak position, flux, and width, we fit these lines with a Gaussian profile, after modelling the local stellar continuum 
with a slope, if present.
We apply this procedure to all emission and absorption lines except for \lya~that we analyse separately due to its peculiar properties (i.e. its resonant nature, see Section~\ref{subsec:lya_fit}).

To estimate the uncertainties on the fit, we perform 1000 Monte Carlo realisations of the spectra. Each realisation is drawn randomly from a Gaussian distribution with mean and variance corresponding to the observed spectrum flux and variance. We then define the uncertainty on the line properties as the half distance between the 16th and 84th percentiles.
In Table~\ref{tab:table_fitlines}, we report the line properties obtained from our fit for all the lines with a SNR $>3$. 
In our error budget, we include systematic uncertainties due to absolute flux calibration of 5\% and 20\% for MUSE and SINFONI data, respectively.

From the wavelength position of the emission lines' peak we estimate the galaxy systemic redshift $z_\textnormal{sys}=3.39535\pm0.00025$.
We limit this approach to emission lines since the interstellar medium (ISM) absorption features appear blueshifted because of outflows, see Section~\ref{subsec:outflows}.

Finally, we measure the rest-frame equivalent width (EW$_0$) of each line as:
\begin{equation}
\label{eq:ew}
    \text{EW}_0~[\text{\AA}] = \frac{1}{1+z_\text{sys}}\int^{\lambda_f}_{\lambda_0}\left(1-\frac{f_\text{line}(\lambda)}{f_\text{con}(\lambda)}\right)d\lambda
\end{equation}
where $\lambda_0$ and $\lambda_f$ are the wavelength limits within which the line fit is performed, and $f_\text{line}$ and $f_\text{con}$ represent the flux density distributions of the line and stellar continuum as a function of the wavelength.
We use a definition of EW$_0$ in which negative values indicate emission while positive values refer to absorption.
Since the optical continuum of the galaxy is not detected in our SINFONI data, we report a 3$\sigma$ upper limit on the flux that, in turn, converts into a 3$\sigma$ lower limit on the line EW$_0$. We estimate $\sigma$ as the median of the error spectrum in the wavelength range within which the line fit is performed.

\subsection{\lya~modelling}
\label{subsec:lya_fit}
Contrarily to the Balmer lines, which escape unobstructed from their production site following recombination, \lya~photons undergo many scattering events.
The number of scatterings depends on the neutral hydrogen column density, geometry and kinematics \cite[see, e.g.][and references therein]{Dijkstra+14}.
Each scattering produces a slight variation in the photon frequency and direction of propagation  \citep{Osterbrock+62}. 
As a consequence of this diffusion process, the spectral characteristics of the emerging radiation encode the properties of the scattering medium along the paths that offered least resistance to the photons \cite[e.g.][]{Dijkstra+16,Gronke+16}.

To adequately model the asymmetric spectral profile of \lya, we resort to Equation~2 by \cite{Shibuya+14b}, i.e.:
\begin{equation}
\label{eq:lya_profile}
    f(\lambda) = A \cdot \exp\left[-\frac{1}{2}\left(\frac{\lambda-\lambda_0^\text{asym}}{\sigma_\text{asym}}\right)^2\right]
\end{equation}
where $A$ is the amplitude and $\lambda_0^\text{asym}$ the peak wavelength of the \lya~line. The asymmetric dispersion, $\sigma_\text{asym}$, is given by $\sigma_\text{asym} = a_\text{asym}\cdot(\lambda-\lambda^\text{asym}_0)+d$, where $a_\text{asym}$ and $d$ are the asymmetric parameter and typical width of the line, respectively. 
An object with a positive (negative) $a_\text{asym}$ value has a skewed line profile with a red (blue) wing. 

Before fitting the \lya~emission, we model the stellar continuum with the \textit{Starburst99}\footnote{\url{https://www.stsci.edu/science/starburst99/docs/default.htm}} synthetic models \citep[][see Section~\ref{subsec:stellar_age_imf} for further details]{Leithere+99}, subtract it from our \lya~spectrum, and apply Equation~\ref{eq:lya_profile} on the residuals.
The \lya~emission is characterised by a prominent redshifted component with a relative velocity (with respect to the systemic redshift) of $403\pm4$~km/s.
We also detected a blue \lya~peak with a flux equal to $\sim$ 5\% of the red component, and a relative velocity of $-294\pm47$~km/s with respect to the systemic redshift. The separation of the blue and red peak is $\Delta v_\text{peak}=697\pm50$~km/s.
This result is in agreement with values reported in \cite{Verhamme+18} for \lya-emitters (LAEs) with a blue peak.

From the fit of the two peaks, we obtain a total \lya~flux of $(1.41\pm0.04)\times10^{-17}$~erg/s/cm$^2$ and an EW$_0=-87\pm10$\AA. 
The equivalent width of \lya\ is larger than the typical values observed in low-redshift LAEs \citep{Rivera-Thorsen+14,Henry+15,Yang+15}, but consistent with other sources at a similar redshift \citep{Erb+14,Trainor+15,Runnholm+20}.
Comparing the peak separation and red peak asymmetry \citep[as first proposed by][]{Verhamme+17,Izotov+18} with the results from \citealt{Kaikiichi+19} (see their Figure~13), we infer that the escape fraction of Lyman continuum photons, $f_\text{esc}^{\text{LyC}}$, is below 15\%.

\begin{figure}
    \centering
    \includegraphics[width=\columnwidth]{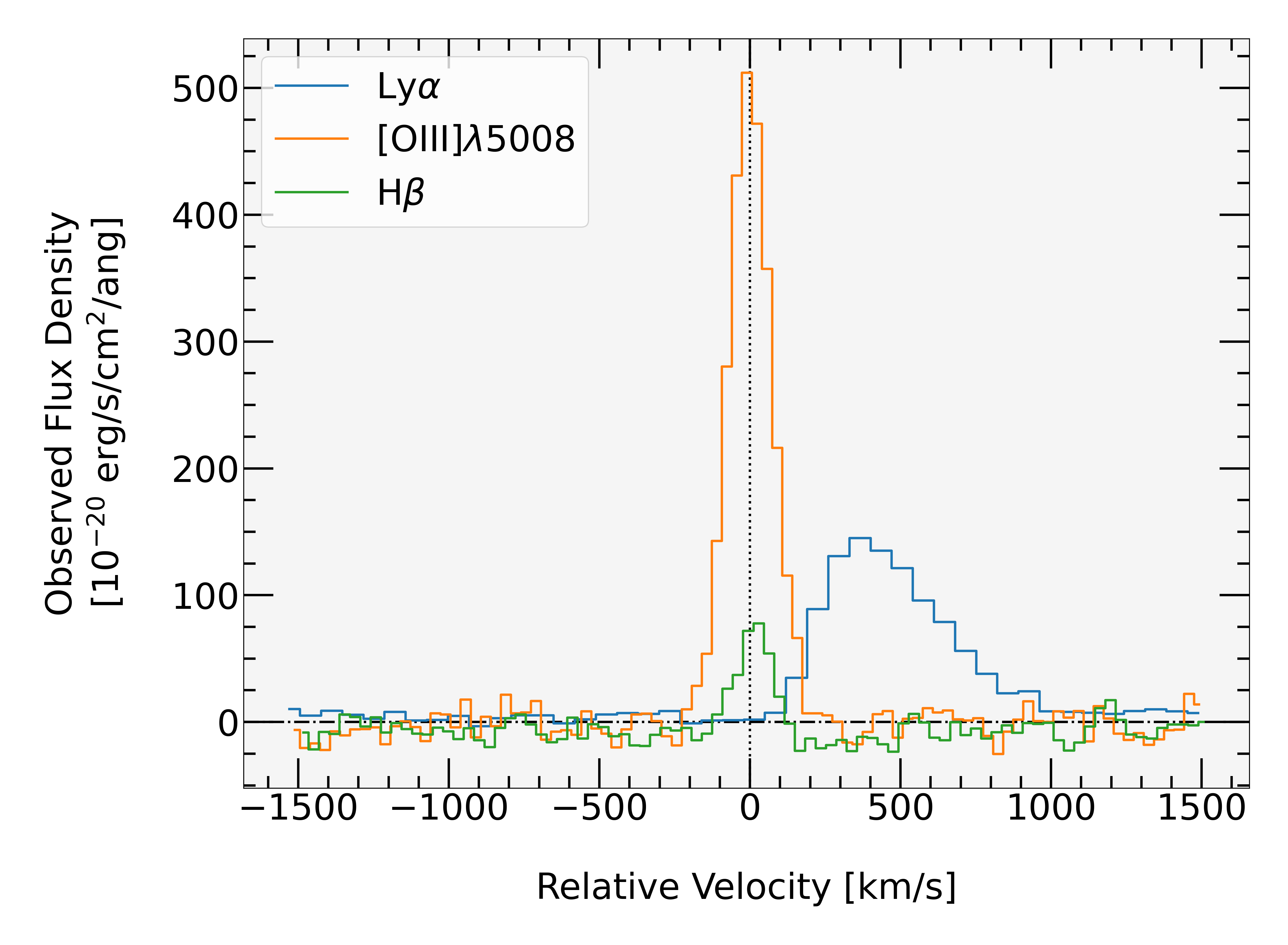}
    \caption{Diagram showing the flux density of the \lya~(blue), \hb~(green) and [OIII]$\lambda5008$ (orange) lines as a function of the line-of-sight velocity (rest-frame).}
    \label{fig:line_velocities}
\end{figure}

If we assume case B recombination and no dust extinction (see Section~\ref{subsec:beta_slope}), we would expect a ratio\footnote{The estimate of the \lya/\hb~intensity ratio has been retrieved by means of the \textsc{Python} package PyNeb by \cite{Luridiana+15} for an electronic temperature $T_e=10^4$~K and density $n_e=10^3$~cm$^{-3}$.} \lya/\hb~$=23.55$.
However, the ratio we measure is of $1.97\pm0.40$, a factor $\sim12$ below the theoretical expectation (for a visual comparison see Figure~\ref{fig:line_velocities}).
This converts into an escape fraction for the \lya~emission of about $f_\text{esc}^{\text{Ly}\alpha}\sim8\%$. 
This value is in good agreement with the global \lya\ escape fraction typically observed at $z\sim 3$ \citep{Gronwall+07,Ouchi+08,Hayes+11}. 
We highlight that the \lya\ spectrum is extracted within the area of the MUSE datacube where we detect the line at a minimum threshold of 2.5$\sigma$. 
This implies that we are neglecting part of the \lya\ at low surface brightness. Hence, our estimate of both the \lya\ flux and $f_\text{esc}^{\text{Ly}\alpha}$ are possibly lower-limits.

As an alternative, the observed discrepancy between the theoretical and observed \lya-\hb~ratio could be ascribed to dust extinction. In this case, the observed ratio could be reconciled with the theoretical expectation by taking into account a colour excess for the nebular emission $E(B-V)_{\rm neb}\simeq0.36$~mag. In the case of dust selective extinction, we would obtain a colour excess for the stellar continuum $E(B-V)_{\rm con}\simeq 0.16$~mag, if we assume a conversion factor $E(B-V)_{\rm con}/E(B-V)_{\rm neb}=0.44$ \cite[see][]{Calzetti+00}. This estimate, however, is not compatible with the observed very steep blue slope of the FUV stellar continuum and the inferred $E(B-V)_{\rm con}=0$~mag (see Section~\ref{subsec:beta_slope}).

A plethora of theoretical studies have demonstrated how the observed \lya~emission profile and its equivalent width depend on the ISM metal and dust content \cite[e.g.][]{Charlot+93}, the relative geometries of the HI and HII regions and the kinematics of the neutral gas \cite[e.g.][]{Neufeld+90,Verhamme+06,Dijkstra+06,Laursen+07}.  
To extract physical information from the \lya~spectral shape, we resort to the commonly used `shell-model' \citep[][]{Ahn+02,Verhamme+06}.
This model consists of a \lya~and continuum emitting source surrounded by a shell of neutral hydrogen, and dust.
It, thus, features four parameters describing the shell: the neutral hydrogen column density of the shell $N_{\rm HI}$, its velocity $v_{\rm exp}$ (defined $>0$ for outflowing), an (effective) temperature $T$ (which also includes the effect of small-scale turbulence), and the dust content -- which
we parametrise as a dust optical depth $\tau_{\rm d}$.
In addition, we use an intrinsic Gaussian emission which we characterise via the
intrinsic \lya~equivalent width EW$_{\rm int}$, and its width
$\sigma_{\rm int}$.

\begin{figure}
    \centering
    \includegraphics[width=.85\columnwidth]{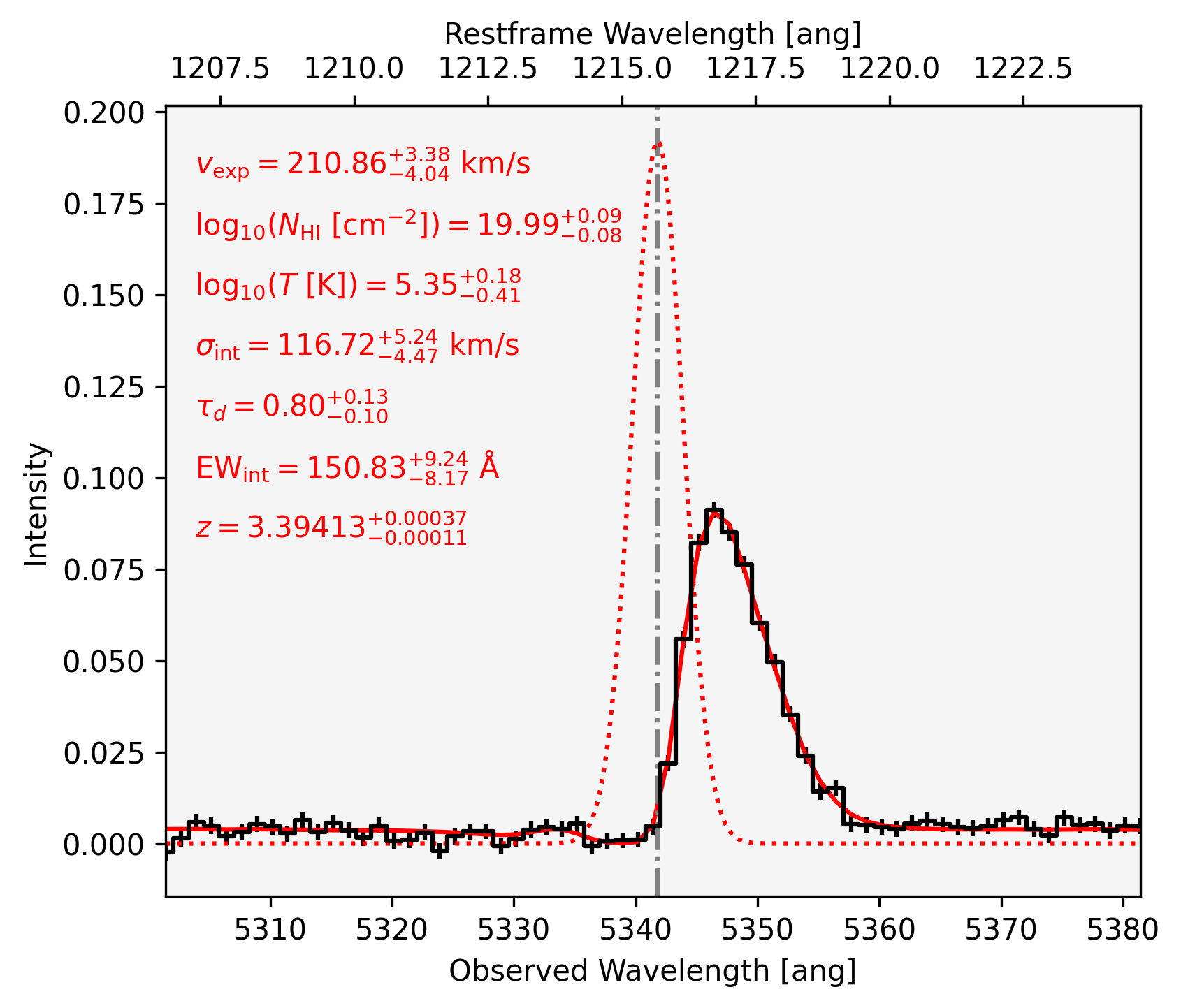}
    \caption{Diagram showing the \lya~intensity spectrum (black solid line) and the best-fit model (red solid line) obtained by means of the radiative transfer code \texttt{tlac} \citep{Gronke+14}. The red dotted line shows the reconstructed shape of the intrinsic \lya~emission, while the vertical dashed-dotted line shows the expected position of the \lya~line according to the systemic redshift.}
    \label{fig:tlac_model}
\end{figure}

To cover this parameter space we specifically employ an improved version of the pipeline described in \cite{Gronke+15} featuring
$12960$ radiative transfer models computed with the radiative transfer code \texttt{tlac} \citep{Gronke+14}.
We carry out the fitting in wavelength space with a Gaussian prior on the redshift $z$.
Furthermore, we smooth the synthetic spectrum by the instrument resolution evaluated at the \lya~observed wavelength \citep[derived from Eq. 8 in][]{Bacon+17}.
We show the result of this fitting procedure in Figure~\ref{fig:tlac_model}.
According to the best-fit model, we derive a $\log_{10}(N_{\rm HI}~[\text{cm}^{-2}])= 19.99\pm 0.09$, $v_{\rm exp} = 211\pm4$ km/s, $\log_{10}(T~[\text{K}]) = 5.35^{+0.18}_{-0.41}$ and $\tau_d = 0.80\pm0.13$. 
While  it is clear that the `shell-model' is an oversimplification of the complex structure and kinematics of \lya~emitting galaxies and their surroundings, it is still unknown how much of the radiative transfer process is captured by the model, and what the fitting parameters physically mean \citep[see discussion, e.g., in ][]{Orlitova+18,Gronke+17,Li+20}. What is clear is that the `shell-model' is able to reproduce the wide range of observed \lya\ spectra well, which may be surprising given its simplicity \citep[see, e.g.,][for an analysis of the fit quality in a large suite of spectra]{Karman+17,Gronke+17}\footnote{Note that while this is a requirement for the reproducibility of the radiative transfer process occurring in nature, it is not trivial to do so -- even with more complex geometries \citep[see discussion of this fact in, e.g.,][]{Gronke+18,Mitchell+20}.}. In addition, the column density $N_{\rm HI}$ as well as the outflow velocity influence the \lya\ spectral shape strongly and are much more robust predictions of the `shell-model' compared to, for instance, the dust optical depth or the effective temperature $T$ (these two parameters typically show large uncertainties and how well they can be tied to their physical counterparts is indeed more uncertain, see \citealt{Verhamme+06,Laursen+09,Gronke+15}). In fact, it has been shown that at least for certain scenarios the outflow velocity and column density of the `shell-model' correlate well with the ones of a more realistic multiphase medium \citep{Gronke+17}. 
In our analysis, we rely on only on these two most robust parameters and we thus conclude that the usage of the `shell-model' to extract physical properties from the observed \lya\ spectrum is well justified. 
We summarise the main results obtained from both the fitting procedure and the analysis of the \lya\ emission in Table~\ref{tab:lya_params}.

\begin{table}
    \centering
    \begin{tabular}{lc}
    \hline
    \hline
    Parameter  & Value \\
    \hline
    $F({\rm Ly}\alpha)$  [erg/s/cm$^2$]   & $(1.41\pm0.04)\times10^{-17}$\\
    EW$_0$ [\AA]                          & $-87 \pm 10$\\
    $\Delta v_{\rm tot}$ [km/s]               & $697\pm50$\\
    $\Delta v_{\rm blue,peak}$ [km/s]         & $-294\pm47$\\
    $\Delta v_{\rm red,peak}$ [km/s]          & $403\pm4$\\
    $f^{{\rm Ly}\alpha}_{\rm esc}$            & $\sim8$\%\\
    $f^{{\rm LyC}}_{\rm esc}$                 & $<15$\%\\
    $\log_{10}(N_{\rm HI} {\rm[cm^{-2}]})$    & $19.99\pm0.09$\\
    $v_{\rm exp}$ [km/s]                      & $211\pm4$\\
    $\log_{10}(T~[\text{K}])$                 & $5.35^{+0.18}_{-0.41}$\\
    $\tau_d$                                  & $0.80\pm0.13$\\
    \hline
    \end{tabular}
    \caption{Table of the results from the line fitting procedure and \texttt{tlac} radiative transfer modelling of the target \lya\ emission, see Section~\ref{subsec:lya_fit}. The line flux $F({\rm Ly}\alpha)$ is corrected for magnification.}
    \label{tab:lya_params}
\end{table}

\begin{table}
    \centering
    \begin{tabular}{ccc}
    \hline
    \hline
    Window  & Wavelength Range \\
    Number  & [\AA]\\
    \hline
    1     & 1268 - 1284\\
    2     & 1309 - 1316\\
    3     & 1360 - 1371\\
    4     & 1407 - 1515\\
    5     & 1562 - 1583\\
    6     & 1677 - 1725\\
    7     & 1760 - 1833\\
    8     & 1866 - 1890\\
    9     & 1930 - 1950\\
    \hline
    \end{tabular}
    \caption{Rest-frame UV spectral windows employed for the measurement of the stellar continuum $\beta$-slope, see Section~\ref{subsec:beta_slope}.}
    \label{tab:spectral_windows}
\end{table}


\begin{figure*}
    \centering
    \includegraphics[width=\textwidth]{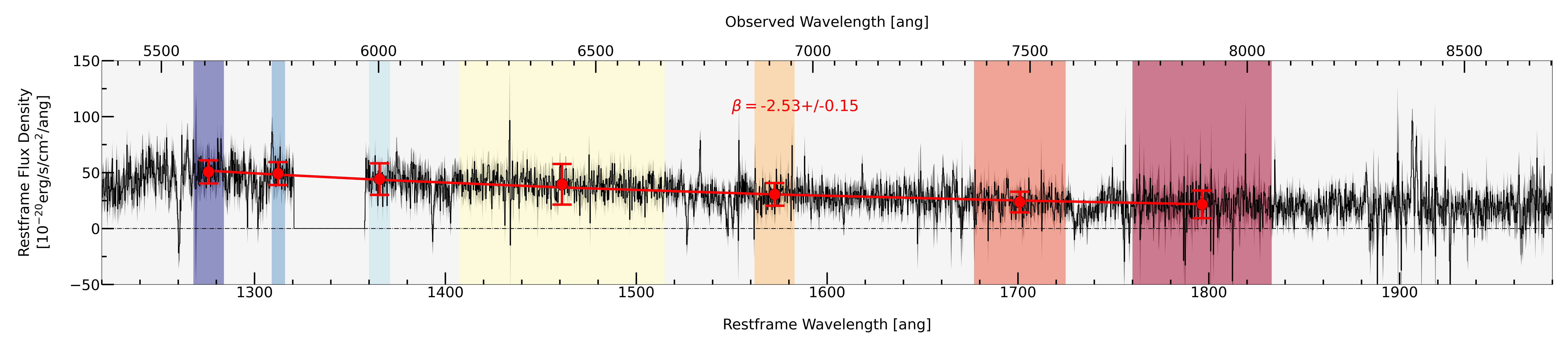}
    \caption{Result of the fitting procedure adopted to evaluate the galaxy $\beta$-slope. With a black solid line, we show the 1220-1980\AA~wavelength region of the galaxy rest-frame UV integrated spectrum. The grey-shaded area around it displays the spectrum $\pm1\sigma$ error. The vertical coloured-shaded areas delimit the 9 spectral regions (see Table~\ref{tab:spectral_windows}) within which we evaluate the galaxy flux (red circles), and its error, we use for the fit. The best-fit power law (red solid line) is presented too.}
    \label{fig:beta_slope}
\end{figure*}

\begin{figure*}
    \centering
    \includegraphics[width=.85\textwidth]{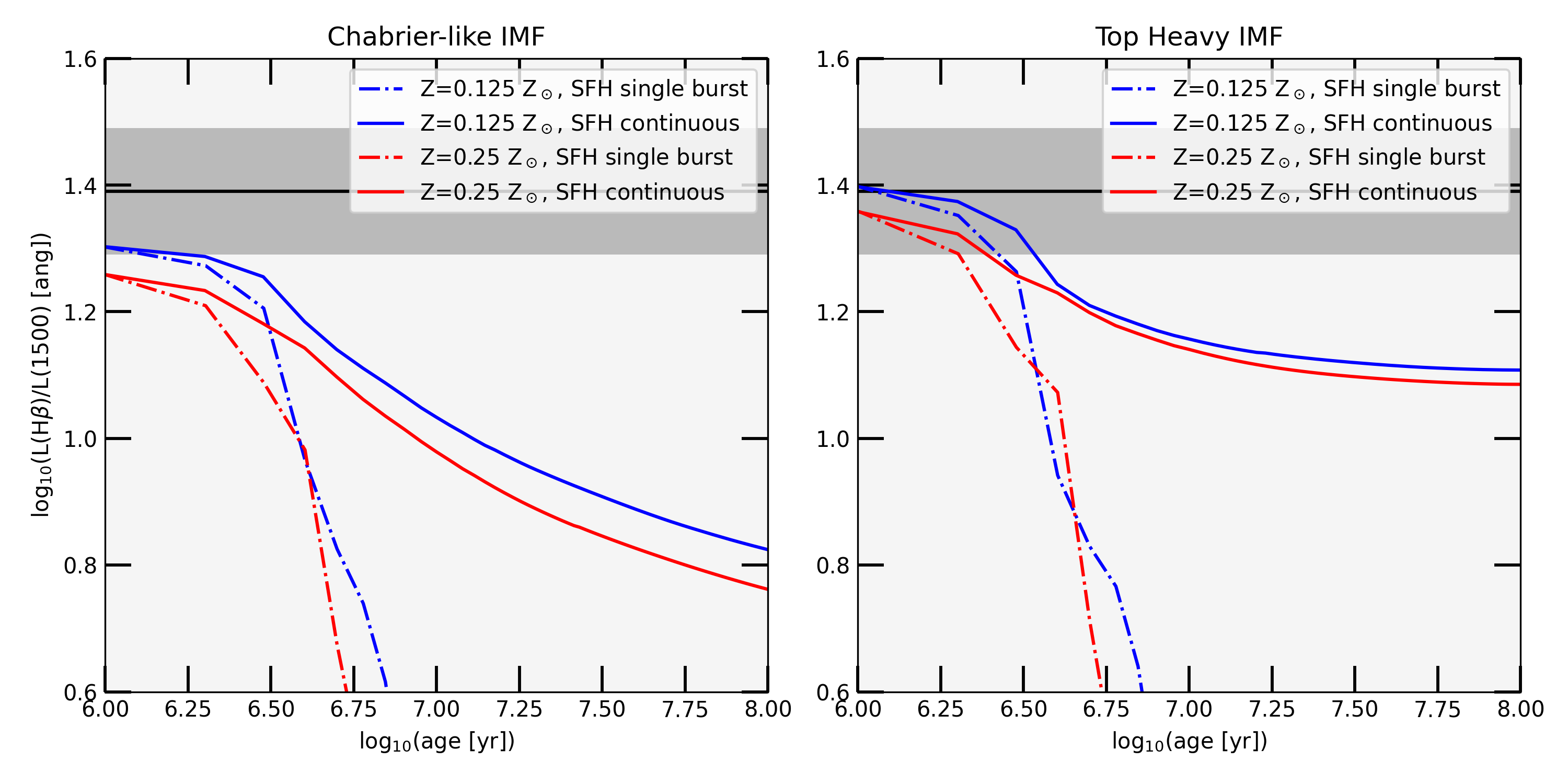}
    \caption{Theoretical tracks of the evolution of the $L(\text{H}\beta)/L_\nu(1500\text{\AA})$ (rest-frame) with stellar population age, and depending on the IMF: Chabrier-like (left panel) or top-heavy (right panel). The tracks are colour-coded according to the stellar metallicity adopted, 0.125 Z$_\odot$ (\textit{blue tracks}) and 0.25 Z$_\odot$ (\textit{red tracks}). Besides, they are reported with a solid line if the SFH used in the modelling is continuous, or with a dashed-dotted line in the case of a single burst model. The tracks have been obtained from the synthetic models of \textit{Starburst99} \citep{Leithere+99}. The horizontal black solid line shows our estimate of the $L(\text{H}\beta)/L_\nu(1500\text{\AA})$ ratio while the grey-shaded area is indicative of its associated error.}
    \label{fig:hb_uv_ratio}
\end{figure*}


\section{Galaxy properties}
\label{sec:galaxy_prop}
In the following Section, we derive the physical properties (e.g. dust extinction, nebular metallicity, star formation rate) of our target while the analysis and interpretation of the results presented in the following will be discussed in the next Section.

\subsection{AGN and SF diagnostics}
\label{subsec:agn_or_sfr}
As a first step in the analysis of our target spectra, we investigate which mechanism is ionising the galaxy ISM, thus driving the emission of the lines.
In particular, we want to understand whether the emission lines that we detect are powered by star-formation only, or if the contribution of an AGN is present. 
From the comparison of the emission line profiles (absence of blue/red wings, broad components) and because of the narrow width of the emission lines ($\leq 200$~km/s, see $\sigma$ values in Table ~\ref{tab:table_fitlines}), it is unlikely that our target hosts an unobscured type-1 AGN \citep[e.g.][]{McCarthy+93,Corbin+96,Humphrey+08,Matsuoka+09}.
The absence of both NV$\lambda1240$ and CIV$\lambda1550$ in emission corroborates this finding. 
Furthermore, when considering UV diagnostic diagrams such as EW([CIII]$\lambda 1907,09$) vs [CIII]$\lambda 1907,09$/HeII$\lambda 1640$ \citep{Nakajima+18}, our target is securely located among the purely star-forming population (e.g. away from the type-2 AGN, composite, and LINERs regions).
Hence, our galaxy appears to be a purely star-forming source.

\subsection{Dust extinction}
\label{subsec:beta_slope}
We estimate the dust extinction affecting the overall galaxy by considering the UV $\beta$-slope.  
As widely implemented in the literature, we fit the observed UV continuum of our target with a power law, expressed as:
\begin{equation}
\label{eq:power_law_fit}
    f(\lambda)~[\text{erg/s/cm$^2$/\AA}] \propto \lambda^\beta
\end{equation}
Similarly to \cite{Calzetti+94}, we define 9 spectral windows in the range 1200-2600\AA\ (see Table~\ref{tab:spectral_windows}) that are carefully designed to remove from the fitting procedure all the relevant absorption features, as well as the MUSE Na Notch filter and strong telluric absorption residuals.
We measure the integrated flux and associated uncertainty of each spectral window and we fit them with Equation~1 from \cite{Castellano+12}, see Figure~\ref{fig:beta_slope}.
The value we obtain from the fit is $\beta=-2.53\pm0.15$, in line with the results found for other low-mass galaxies at similar redshift \cite[][]{Castellano+12,Bouwens+16,Vanzella+18}.
Such low value of the $\beta$ parameter is typical of stars with steep blue UV slopes, i.e. young and unobscured stellar populations.
Indeed, if we convert the measured $\beta$ into the colour excess of the stellar continuum $E(B-V)_\text{con}$ via the relation by \cite{Meurer+99}, we obtain a value that is compatible with $E(B-V)_\text{con} = 0$~mag only within the error bar ($E(B-V)_\text{con} = -0.08\pm0.08$~mag).
Despite the fact that the $\beta-E(B-V)_{\rm con}$ relation depends on metallicity and star formation history \cite[e.g.][]{Kong+04, Dale+09, Munoz-Mateos+09, Reddy+10, Reddy+18, Schaerer+13, Zeimann+15}, as well as on stellar mass and age \cite[e.g.][]{Buat+12, Zeimann+15, Bouwens+16a}, we consider our estimate robust. In fact, if we assume an $E(B-V)_{\rm con}>0$~mag (e.g. $E(B-V)_{\rm con}=0.16$~mag from the \lya-\hb\ ratio, see Section~\ref{subsec:lya_fit}), we would obtain a $\beta$-slope corrected for dust extinction even more extreme (e.g. $\beta=-3.43\pm0.14$) and hardly reconcilable with any known physical scenario. 
Finally, a preliminary analysis of our target far-infrared continuum emission (ALMA observations, PI: E. Iani, Zanella et al. \textit{in prep.}) further corroborates our finding.


\subsection{Nebular metallicity}
\label{subsec:metallicity}
Thanks to the variety of ISM emission lines we detect in the galaxy FUV spectrum, we can estimate the nebular metallicity of our target by considering the He2 -- O3C3 diagnostic diagram by \cite{Byler+20}. 
Through Equation 8 by \cite{Byler+20}, we measure a metallicity $12+\log_{10}(O/H)=7.94\pm0.07$, that corresponds to $\text{Z}=0.18\pm0.04$~Z$_\odot$, if we assume the solar value of $12+\log_{10}(O/H)=8.69\pm0.05$ \citep{APrieto+01}.
From the comparison with the model grids, we can infer a rough estimate for the ionisation parameter (U) of $\log_{10}(\text{U})\sim-2$. 

An independent estimate of the gas-phase metallicity can be derived also by considering the [OIII]$\lambda5008$/\hb~ratio \citep{Maiolino+08}. In this case, we obtain $12+\log_{10}(O/H)\sim7.89$ that corresponds to $\text{Z}\sim0.16$~Z$_\odot$.
Even though this last estimator has been proven to be strongly dependent on the ionisation parameter \citep[e.g.][]{Kewley+19_rev}, the measurement is in good agreement with the He2 -- O3C3 estimate. 

\subsection{Star formation rate}
\label{subsec:sfr}
We estimate the star formation rate (SFR) of our target in two ways: from the \hb~luminosity, and from the luminosity of the UV continuum at 1500\AA.
In both cases, we apply the recipes by \cite{Kennicutt+98} after correcting them for a Chabrier IMF\footnote{To transform from a Salpeter to a Chabrier IMF, the derived SFR has to be divided by a 1.7 factor.} (the original relations being defined for a Salpeter IMF, \citealt{Salpeter+55}).

To convert the H$\beta$ luminosity into SFR, we use the relation:
\begin{equation}
\label{eq:sfr_ha}
    \text{SFR(H}\beta)~[\text{M}_\odot/\text{yr}] = 1.33\times10^{-41} L(\text{H}\beta) [\text{erg/s}]
\end{equation}
valid for an electronic temperature T$_e=10^4$K, 
and case B recombination \citep{Osterbrock+06}, 
i.e. all the ionising photons are processed by the gas ($f_\text{esc}^\text{LyC}=0$).
From the above equation, we derive $\text{SFR(H}\beta)=9.9\pm2.3~\text{M}_\odot/\text{yr}$.

Similarly, we can convert the rest-frame UV stellar continuum luminosity at 1500\AA, $L_\nu(1500\text{\AA}$, into SFR via the equation:
\begin{equation}
\label{eq:sfr_continuum}
    \text{SFR}(1500\text{\AA})~[\text{M}_\odot/\text{yr}] = 8.24\times10^{-29} L_\nu(1500\text{\AA}) [\text{erg/s/Hz}]
\end{equation}
where $L_\nu(1500\text{\AA})=(2.30\pm 0.12)\times 10^{28} \text{erg/s/Hz}$ and was extrapolated from the fit of the UV continuum with a power-law, see Section~\ref{subsec:beta_slope}.
From the above equation, we obtain a $\text{SFR(1500\AA)}=1.9\pm0.7~\text{M}_\odot/\text{yr}$.

According to our measurements, the ratio between the SFR(\hb) and SFR(1500\AA) is equal to $5.2\pm2.3$. 
This discrepancy is well-expected since Hydrogen lines and the UV stellar continuum trace the current star formation of a galaxy over different timescales. In fact, while Balmer lines allow to derive the {\it instantaneous} star formation, i.e. star formation of the last $\lesssim 10$~Myr, the UV -- SFR relation implicitly assumes a continuous and well-behaved star formation history, ongoing for at least $100$ Myr. 
To properly account for this difference in timescales, the multiplicative factor in Equation~\ref{eq:sfr_continuum} has to be corrected.
In particular, if the star formation timescale is $\lesssim 10$ Myr, the SFR(1500\AA) can be underestimated up to a factor $\sim 3.5$ \citep[e.g.][]{Calzetti+13}.
With this correction, the two estimates agree within the errors.

It is useful to notice that discrepancies between the SFR estimators presented above can give insights on the timescale over which star formation processes are taking place, and hence, on the age of the youngest stellar populations (see Section~\ref{subsec:stellar_age_imf} for further details).

In the following, unless differently stated, we assume as SFR the one obtained from Equation~\ref{eq:sfr_ha}, i.e. $\text{SFR(H}\beta)=9.9\pm2.3\text{M}_\odot/\text{yr}$.

\subsection{Stellar age and IMF}
\label{subsec:stellar_age_imf}
The ratio between the de-reddened \hb~luminosity and the $L_\nu(1500\text{\AA})$ gives an estimate of the stellar population age, as this ratio decreases with increasing stellar age. 
In the left panel of Figure~\ref{fig:hb_uv_ratio}, we present the $L(\text{H}\beta)$/$L_\nu(1500\text{\AA})$ evolution with time, assuming a Chabrier-like IMF\footnote{The Chabrier-like IMF we adopt in this paper is:
\begin{equation*}
  \xi(\text{M}) \propto
    \begin{cases}
      \text{M}^{-1.3} & \text{if $0.1\leq \text{M}~[\text{M}_\odot] < 1$}\\
      \text{M}^{-2.3} & \text{if $ 1\leq \text{M}~[\text{M}_\odot] \leq 100$}
    \end{cases}       
\end{equation*}
}, different stellar metallicities (0.125~Z$_\odot$ and 0.25~Z$_\odot$), and different star formation histories (SFH, single burst and continuous star formation).
To construct the $L(\text{H}\beta)/L_\nu(1500\text{\AA})$ tracks, we use the spectrophotmetric synthetic models of \textit{Starburst99}.
Among the available discrete tracks, we choose those with stellar metallicity matching the nebular metallicity that we measured through the He2 -- O3C3 diagnostic diagram, see Section~\ref{subsec:metallicity}. 
We expect the stellar metallicity to be comparable or lower than the metallicity of the ISM from which stars form.
About the SFH, the single burst and continuous star formation depict extreme and opposite scenarios making the tracks of the real galaxy SFH possibly an intermediate solution between the two.
From the comparison of the theoretical tracks with our measurement, we conclude that our target hosts a stellar population younger than 10~Myr.

The $L(\text{H}\beta)/L_\nu(1500\text{\AA})$ that we measure is high, at the limit of the ratios predicted by \textit{Starburst99} for a Chabrier-like IMF. We investigate whether the assumption of a more exotic solution (e.g. top-heavy IMF) could alleviate the tension between observations and models. In the right panel of Figure~\ref{fig:hb_uv_ratio} we show that a stellar population with a top-heavy IMF\footnote{The top-heavy IMF we adopt in this paper is \cite[][]{Zanella+15}:
\begin{equation*}
  \xi(\text{M}) \propto 
      \text{M}^{-1.35} ~~~ \text{if $5\leq \text{M}~[\text{M}_\odot] \leq 100$}
\end{equation*}
} would show higher $L(\text{H}\beta)/L_\nu(1500\text{\AA})$ values than the Chabrier-like IMF case, and seems to be more compatible with our observations.

\begin{figure}
    \centering
    \includegraphics[width=\columnwidth]{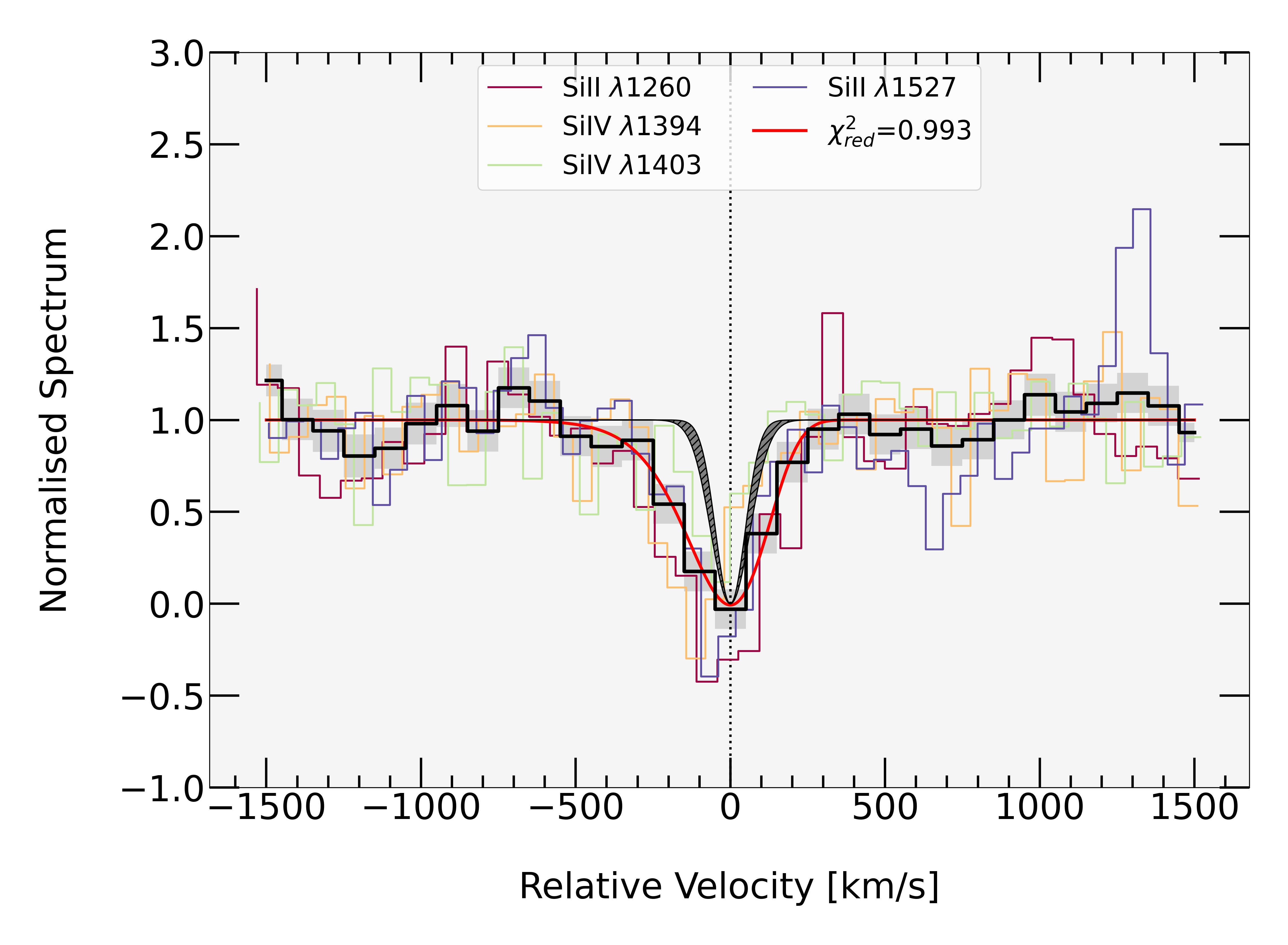}
    \caption{Diagram showing the normalised profile of the strongest FUV absorption lines as a function of the line-of-sight velocity (rest-frame). The black solid line shows the mean profile of the absorptions, while the grey shaded area is indicative of the $1\sigma$ uncertainty. The grey hatched Gaussian represents the range of resolution in velocity of MUSE obtained from Equation 8 in \protect\cite{Bacon+17}. The red solid line shows the best-fit of the stacked spectrum performed with a skewed gaussian model.}
    \label{fig:line_abs_velocities}
\end{figure}

\begin{figure}
    \centering
    \includegraphics[width=\columnwidth]{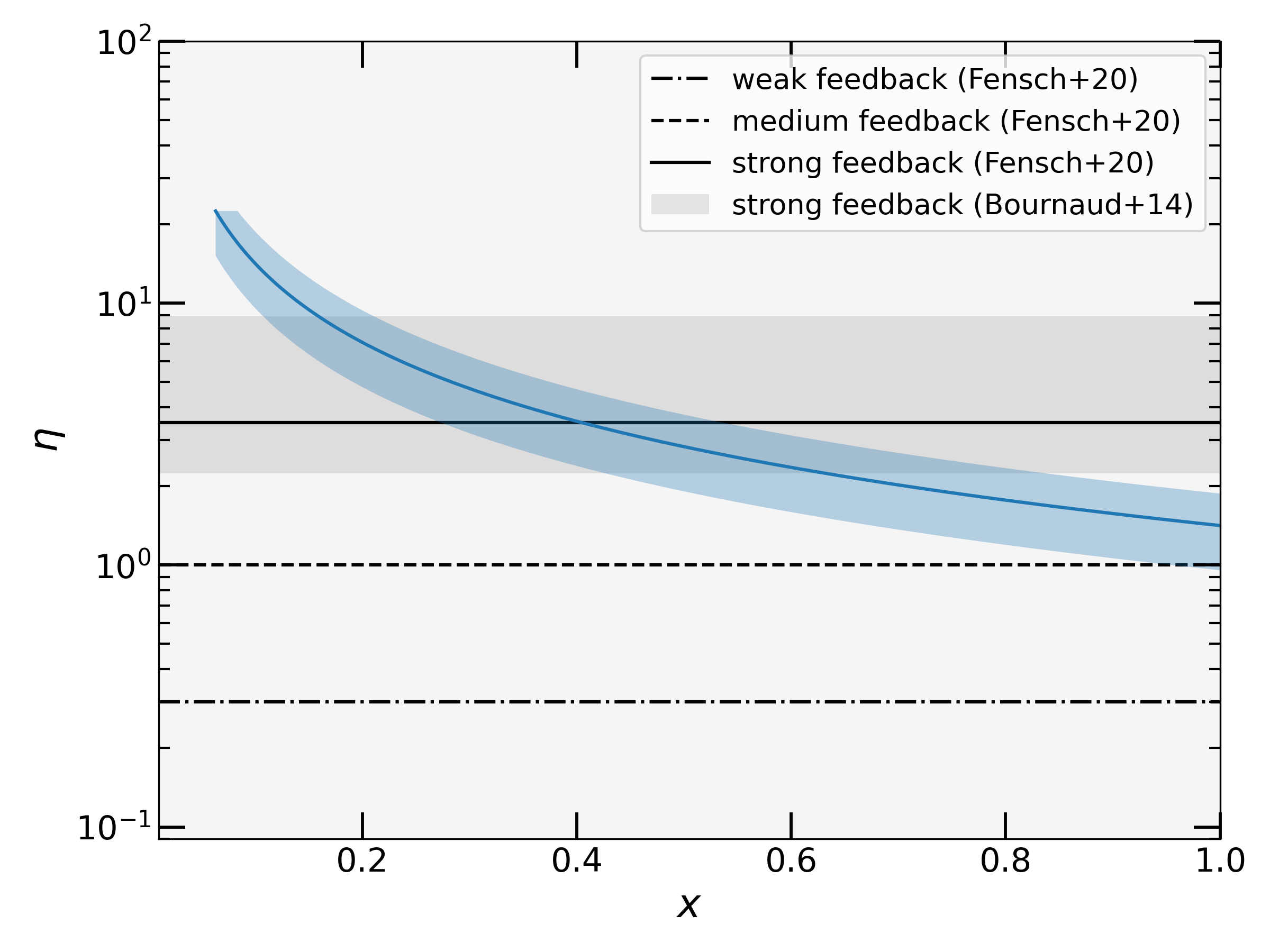}
    \caption{Diagram of the mass-loading factor $\eta$ as a function of $x$. The horizontal black lines show the average values of galaxy-wide $\eta$ that have been found in the simulations by \protect\cite{Fensch+20} implementing weak ($\eta=0.3$), medium ($\eta=1$) and strong ($\eta=3.5$) stellar feedback calibrations, respectively. The horizontal grey shaded area shows the range of $\eta$ that can be obtain only by hydro-dynamical simulations implementing recipes of strong supernovae feedback (e.g. G1, G2 and G3 models) in \protect\cite{Bournaud+14}.}
    \label{fig:outflow_geometries}
\end{figure}

\begin{figure}
    \centering
    \includegraphics[width=\columnwidth]{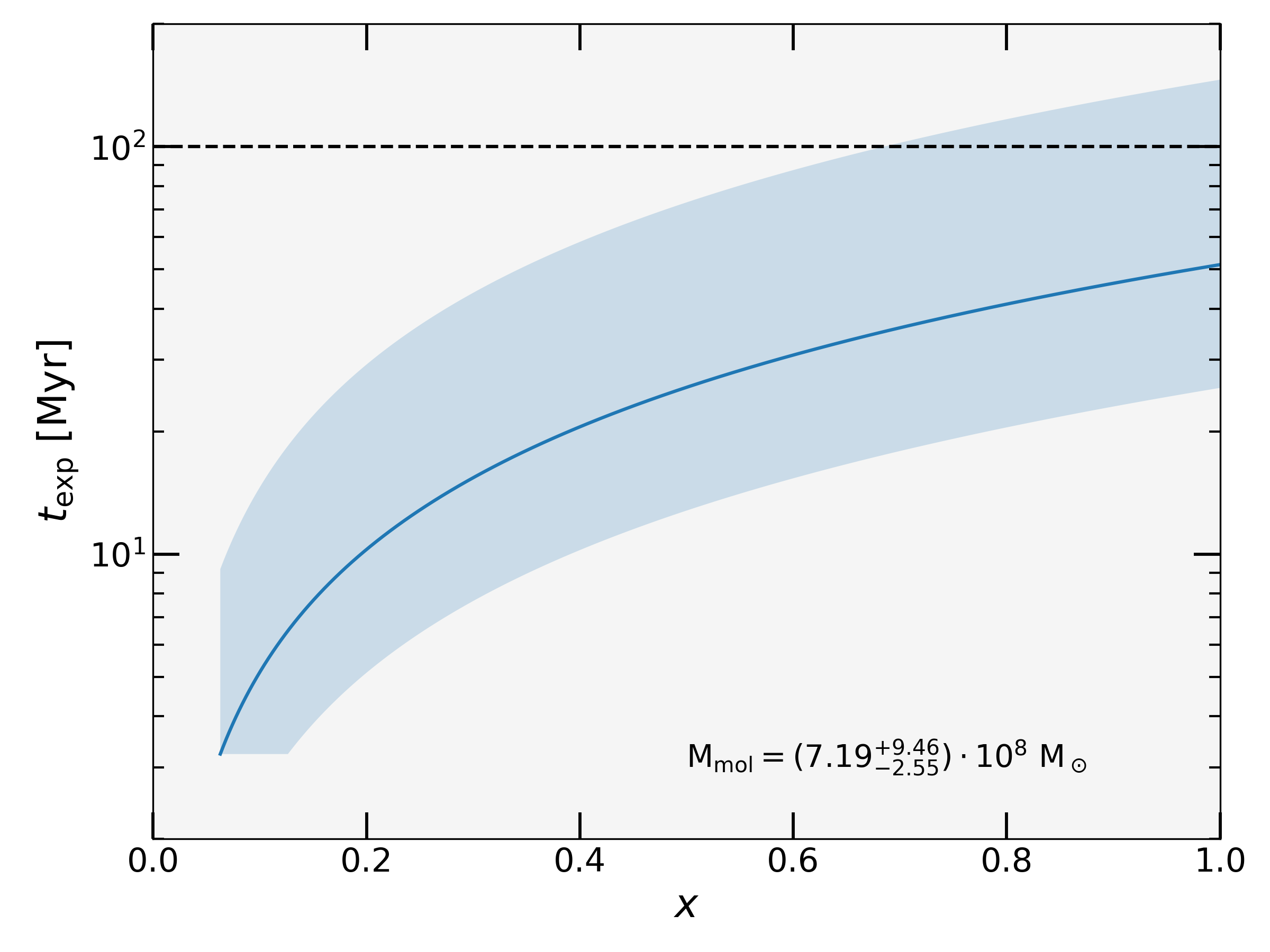}
    \caption{Diagram of the timescale of survival to star formation feedback of clumps ($t_\text{exp}$) as a function of the fraction of HI in the outflow $x$.}
    \label{fig:clumps_timescale_starburst}
\end{figure}


\section{Discussion}
\label{sec:discussion}
According to the results presented in the previous Sections, A2895a is a lensed star-forming \lya-emitter at $z\simeq 3.4$ that hosts four compact clumps. 
Similarly to what typically found in LAEs \cite[e.g.][ and references therein]{Ouchi+20}, the galaxy has a low metallicity ISM (${\rm Z}\simeq0.2~{\rm Z}_\odot$) and a blue FUV stellar continuum ($\beta\simeq-2.5$) which implies a stellar extinction $E(B-V)_{\rm con}\sim 0$.

The \hst\ PSF sets an upper-limit on the clumps' size of 280 pc.
Based on the clumps' size -- stellar mass relation by \cite{Cava+18}, we can associate to the individual clumps an upper-limit on their stellar mass of $2\times10^8~{\rm M}_\odot$.

The clumps contribute to $\sim 60$\% of the galaxy FUV emission that appears to be powered by a young stellar population of hot and massive stars with an age of less than 10~Myr, as obtained from the $L(\text{H}\beta)/L_\nu(1500\text{\AA})$ ratio \citep[e.g.][]{Leithere+99,Zanella+15} and in agreement with studies of LAEs \citep[e.g.][]{Nakajima+12}. 
Despite the fact that the emission lines (but the \lya) detected in the target's FUV and optical spectra are spatially coincident with the FUV continuum probed by \hst, we cannot determine weather the lines arise within the single clumps or from the overall galaxy because of MUSE and SINFONI coarser spatial resolution. 
Yet, several studies have shown that the lines emission predominantly originates from the clumps if their age is $\lesssim 10$~Myr, as in the case of our target \citep{Genzel+11,ForsterS+11a,Zanella+15,Zanella+19}. 
Hence, if we estimate the galaxy SFR from the conversion of the \hb~luminosity, we can assume that the galaxy star formation activity is mainly taking place within the clumps at an overall estimated rate of $\sim 10$~\msunyr.
This sets a lower-limit on clumps sSFR of $1.25\times 10^{-8}~{\rm yr}^{-1}$ that is consistent with the sSFR estimates of compact clumps in \cite{Zanella+19} and that suggests that the detected clumps are forming stars in a `starbursting mode' \citep[e.g.][]{Zanella+15,Bournaud+15}.   

Finally, The $L(\text{H}\beta)/L_\nu(1500\text{\AA})$ ratio hints to a star formation activity that follows a top-heavy IMF. 
A similar result was already obtained for another very young clump (age $\lesssim 10$ Myr) hosted by a $z \sim 2$ galaxy \citep{Zanella+15}. Yet, an analysis on a statistical sample is needed to draw more robust conclusions in this regard.

\subsection{ISM outflows}
\label{subsec:outflows}
The FUV absorption lines have larger velocity dispersion ($\sigma \sim 90$~km/s) than the emission lines ($\sigma \sim 40$~km/s, see Table~\ref{tab:table_fitlines}).
Besides, the absorption features display an asymmetrical profile skewed towards shorter wavelengths, a blue wing, that becomes particularly evident when stacking the absorption lines together (e.g. SiII $\lambda1260,1527$ and SiIV $\lambda1394,1403$), see Figure~\ref{fig:line_abs_velocities}.
Both the larger velocity dispersion and the presence of blue wings in UV absorption lines are typically ascribed to gas outflows in the galaxies' ISM \citep[e.g.][]{Pettini+00,Quider+09,DessaugesZavadsky+10,Erb+12,Patricio+16}.
This conclusion is also supported by our analysis of the \lya\ spectral shape according to which the \lya\ photons are propagating within a medium that is expanding at a velocity of $v_{\rm exp}= 211\pm 4$~km/s, see Section~\ref{subsec:lya_fit}.

Independently of the \lya\ modelling, the analysis of the observed UV absorption line profiles is often used to infer the maximum velocity of galactic outflows. One way to achieve this result is by means of the $v_{90}$ parameter \cite[][]{Prochaska+97,Wolfe+98}, i.e. the blue-shift velocity where the lines' wing intensity reaches 90\% of the continuum intensity.
To estimate $v_{90}$ from our FUV spectrum, we first fit the normalised and stacked absorption line profile with a skewed Gaussian. From the best-fit model, we derive a maximum outflow velocity of $-363\pm53$~km/s. We estimate the error following a MonteCarlo procedure, i.e. by perturbing the normalised stacked spectrum according to its associated error 5000 times and measuring the half distance between the 16th and 84th percentiles of the output $v_{90}$ distribution. 
Given the galaxy SFR, the maximum outflow velocity we derived is in good agreement with what has been observed in other galaxies at lower redshifts \cite[i.e.][]{Chisholm+15, Heckman+15, Bordoloi+16}.

The value of $v_{90}$ is an independent estimate of the outflow velocity to the one obtained from the modelling of the \lya\ emission $v_{\rm exp}$. 
Comparing the two estimates, we obtain $v_{90}>v_{\rm exp}$.
This is due to the fact that while $v_{90}$ is indicative of the maximum velocity of the outflow, the \lya\ photons are likely susceptible to a mean (e.g., mass weighted) outflow velocity.
We underline that this result is not affected by the geometry and inclination of the outflow. If we assume that the outflow is not spherical, regardless of the inclination of the galaxy, our $v_{90}$ estimate would represent a lower limit. In fact, we would be measuring only the outflow component projected along the observer line-of-sight. On the contrary, $v_{\rm exp}$ does not suffer from projection effects as also photons initially escaping along a path different to the line-of-sight can be scattered back into the observer’s direction. 
Therefore, even though the maximum outflow velocity could significantly increase, this would not create any tension with the actual velocity estimate derived by the \lya\ modelling.


\subsection{Star formation feedback and outflows energetics}
\label{subsec:mass_loss}
If we assume that the detected ISM outflows are the direct consequence of star formation feedback taking place only within the four star-forming regions harboured in our target, we can estimate the rate at which star formation expels the ISM from the four clumps, i.e. the gas mass-loss rate $\dot{M}$, as \cite[following][]{Pettini+00}:
\begin{equation}
\label{eq:final_mass_loss_corpus}
    \dot{M}~[\text{M}_\odot/\text{yr}] = \frac{3.09\times 10^{-22}}{x} \cdot \left( \frac{r}{[\text{kpc}]} \right) 
    \cdot \left( \frac{N_\text{HI}}{[\text{cm}^{-2}]} \right) \cdot \left( \frac{v_\text{exp}}{[\text{km/s}]} \right) 
\end{equation}
where $N_{\rm HI}$ is the Hydrogen column density, $v_{\rm exp}$ the expansion velocity of the outflow,  $r$ is the radius of the expanding shell, and $x$ is the ratio between the mass of the Hydrogen atom $m_{\rm HI}$ and the average particle mass in the outflowing medium $m_p$ (i.e. $x=m_{\rm HI}/m_p$).
For a complete description on how Equation~\ref{eq:final_mass_loss_corpus} has been derived see Appendix~\ref{appendix:deriving_mass_loading_factor}. 

For both $N_{\rm HI}$ and $v_{\rm exp}$, we adopt the values derived from the analysis of the \lya\ emission, i.e. $\log_{10}(N_{\rm HI}~{\rm [cm^{-2}]})=19.99\pm0.09$ and $v_{\rm exp}=211\pm4$~km/s (see Section~\ref{subsec:lya_fit}).
We highlight that the parameters derived from the modelling of the \lya\ spectrum probe the galaxy medium along the so-called \textit{path of least resistance}, i.e. the path with the lowest optical depth along which the \lya\ photons diffuse out as a consequence of resonant scattering \citep{Dijkstra+16, Eide+18}. Therefore, by inserting these parameters into Equation~\ref{eq:final_mass_loss_corpus}, we implicitly assume that the \lya\ probes the wind medium since the outflowing material is likely to have a significantly lowered optical depth (and possibly also lower column density, \citealt{Behrens+14}). 

We infer the radius of the expanding shell $r$ as given by the product between the gas expansion velocity $v_{\rm exp}$ and the age of the clumps' stellar population (10~Myr, see Section~\ref{subsec:stellar_age_imf}), i.e. $r=v_{\rm exp}\cdot t_{\rm age}=2.2\pm0.2$~kpc. This assumes that the outflows were in place since the beginning of the on-going burst of star formation and kept a constant expansion velocity through time. Even with these simplifying assumptions, we find that $r$ encompasses most of the observed \lya\ emission of our target from which we derive the $v_{\rm exp}$ and $N_{\rm HI}$ parameters.

For the HI fraction of the outflowing medium $x$, previous studies have often adopted $x=1$ \citep[e.g.][]{Pettini+00,Verhamme+08}, thus considering that the outflowing material consist of HI only. To take into account the possible presence of heavier elements, a few studies have lowered the estimate of $x$ to 0.74 based on the fact that the ISM of galaxies is mainly a mixture of HI (90\% of the total ISM mass) and atomic Helium (10\%), while the other metals contribute less than 0.1\% \cite[e.g.][]{Genzel+08}.
However, star forming regions are rich in molecular gas (mostly H$_2$) and spatially resolved studies of galaxies in the local Universe have shown that the molecular phase of outflows constitutes a significant amount of the ejected material \citep[e.g.][]{Weiss+99,Walter+02,Sakamoto+06,Bolatto+13}.
In particular, \cite{Smirnova+17} found that in star-forming regions of galaxies in the local Universe the mass of $H_2$ and HI are comparable. 
According to this finding, $x=0.67$.
Because of the uncertainties related to the above assumptions (mainly on the metals and $H_2$ content), in the following we assume as reference value for $x$ the interval 0.6-0.8.

Finally, Equation~\ref{eq:final_mass_loss_corpus} is valid if we assume an outflow geometry given by a thin spherical expanding shell. 
Despite the fact that a few observational and theoretical works have shown evidence that \lya~photons scatter off a bipolar outflow \cite[e.g.][]{Blandford+74,Suchkov+94,Duval+16}, we assume a thin spherical expanding shell since we do not have any direct evidence pointing to a bipolar geometry. 
We also highlight that $\dot{M}$ gives an estimate of the overall mass-loss rate of the four detected clumps, and that each clump could be characterised by a bipolar outflow expanding in a different direction and with a different opening angle. We however report the effects of alternative geometries, i.e. biconical and double spherical sector outflows, in Appendix~\ref{appendix:outflow_geometries}.

In Figure~\ref{fig:outflow_geometries}, we present a track for the mass loss rate normalised by the galaxy SFR (i.e. $\dot{M}/\text{SFR}$), the so-called mass loading factor $\eta$. This estimate can be considered as the average mass-loading factor of the clumps, if we assume that the estimated SFR and $\dot{M}$ are equally distributed among the four detected clumps. 
The SFR we use for this estimate is the value obtained from the conversion of the \hb\ luminosity, i.e. ${\rm SFR(H\beta)} = 9.9\pm2.3$~\msunyr.
Following \cite{Swinbank+07}, we limit the track in Figure~\ref{fig:outflow_geometries} to the minimum value of $x$ for which outflows are feasible, i.e. $x=0.06$ and a corresponding value of $\eta\simeq22$.   
Independently on $x$, $\eta>1$. In particular, for $x$ in between 0.6 and 0.8 we obtain $\eta=2.4$ and $\eta=1.8$, respectively. 
These mass loading factors are in good agreement with those found by \cite{Genzel+11} who analysed the spectral profile of optical emission lines (H$\alpha$ and [OIII]) from massive clumps ($10^9-10^{10}\text{M}_\odot$) characterised by high-velocity outflows ($350-1000$ km/s) in five star-forming galaxies at $z\sim2$, and found $\eta$ ranging from $1-9$.
Similar mass loading factors ($\eta=2-9$) were also found in \cite{Newman+12a}.

We also compare our findings with results obtained from hydro-dynamical simulations of $z \sim 2$ clumpy galaxies by \cite{Bournaud+14} and \cite{Fensch+20}. 
In their study, \cite{Bournaud+14} found evidence that gas clouds with masses of a few $10^7\text{M}_\odot$ are rapidly blown up by star formation feedback, while massive clumps ($\approx 10^8\text{M}_\odot$) are long-lived and have lifetimes that range from $200-700$ Myr.
For such massive clumps, \cite{Bournaud+14} found that the mass loading factor of the clumps that formed in simulations implementing strong SNe feedback (i.e. simulations G1, G2 and G3) follows a distribution that has mean value of 1.6 and a tail that extends up to 10 (see their Figure~9, left panel).
Such high values were hardly recovered in the case of simulations with a weaker SNe feedback (e.g. G'2 model) that have mass loading factors in the range 0.1 - 5 with a median value of 0.7. 
A similar result was recently obtained by \cite{Fensch+20}, who found that the average mass loading factor in simulations of galaxies at $1<z<3$ hosting clumps with average stellar masses of $10^8\ {\rm M}_\odot$ implementing strong SNe feedback is of 3.5, independently from the galaxy gas mass fraction (see their Table~3).
Lower values of $\eta$ where found only for weak ($\eta=0.3$) and medium ($\eta=1$) stellar feedback, i.e. simulations where the energy from type-II supernovae is mostly ($\geq 90$\%) released thermally and not in kinetic form.
Comparing the results by \cite{Bournaud+14} and \cite{Fensch+20} with our findings, the values of $\eta$ we infer seem to be consistent with the simulations implementing a strong/medium SNe feedback. 

Knowing the gas mass-loss rate, we can estimate the timescale needed for the stellar feedback to expel the gas from the clumps, thus quenching their star formation activity. 
We derive this quantity in the case of a `semi-closed box' model, i.e. neglecting the possible presence of inflowing gas that could replenish the reservoir of clumps and therefore sustain star formation for a longer period \cite[e.g.][]{Dekel+13,Bournaud+16,Fensch+20}. 
Given this assumption, we derive a lower-limit on the gas removal timescale $t_\text{exp}$ that is given by $t_\text{exp} = \text{M}_\text{mol}/\dot{M}$,
where $\text{M}_\text{mol}$ is the clumps' molecular gas mass.
We estimate $\text{M}_\text{mol}$ by considering the integrated Schmidt-Kennicutt relation reported by \cite{Sargent+14}.
In particular, according to our findings on the clumps' sSFR and supported by recent studies targeting young clumps \citep[e.g.][]{Guo+12,Wuyts+12,Wuyts+13,Bournaud+15,Zanella+15,Mieda+16,Cibinel+17,Zanella+19}, we assume that our clumps form stars in a starbursting mode\footnote{For the sake of completeness, we report in Appendix~\ref{appendix:clumps_timescale_ms} the dependence of the gas removal timescale on $x$ if the clumps form stars in a `main-sequence' mode \cite[from the stellar mass -- SFR relation, e.g.][]{Elbaz+07,Rodighiero+11,Whitaker+12,Sargent+14}.}.
In this case, the amount of molecular gas locked into the clumps would be M$_\text{mol} = (7.19^{+9.46}_{-2.55})\times 10^{8}\ \text{M}_\odot$.

In Figure~\ref{fig:clumps_timescale_starburst}, we present the dependence of $t_\text{exp}$ from $x$. Also in this case we limit the track to the minimum value of $x=0.06$.
Independently on $x$, $t_\text{exp}$ is always below $100$~Myr.
In particular, assuming the estimates of $\dot{M}$ presented above, $t_\text{exp}$ ranges between $20-50$~Myr. 
According to these values, the detected clumps would expel their gas on a very short timescale thus stopping their star formation activity in a few tens of Myr.

\section{Conclusions}
\label{sec:conclusions}
In this paper, we have examined the physical properties of a triply-imaged line-emitting galaxy at redshift $z\simeq3.4$ and withdrawn from the sample of lensed clumpy galaxies by \cite{Livermore+15}.
Thanks to our analysis of integral-field spectroscopic data from VLT/MUSE and SINFONI, as well as \textit{HST} rest-frame FUV imaging, we found that:

\begin{itemize}
\item The three multiple images of the galaxy show an irregular FUV morphology that is constituted by four compact clumps whose light accounts for $\sim60$\% of the total galaxy FUV emission and that have sizes $\lesssim 280$~pc and stellar masses $\lesssim 2\times10^8\ {\rm M}_\odot$.

\item The galaxy FUV and optical spectra feature a wide variety of lines both in emission (the brightest are \lya, \hb, and [OIII]$\lambda4959,5008$) and absorption (e.g. SiII, SiIII, SiIV, as well as other fainter metal lines such as O, Al, Fe).
The absorption lines have a wider velocity dispersion ($\sigma \sim 90$km/s) if compared to the FUV and optical emission lines ($\sigma\sim40$km/s).
This suggests that the galaxy ISM is characterised by the presence of outflows. 
From the stacking of all absorption lines, we recover a mild blue asymmetry consistent with outflows with a terminal velocity $\lesssim 350$~km/s.

\item Our target is a star-forming galaxy.  
The blue slope of the FUV stellar continuum ($\beta=-2.51\pm0.12$), the relatively low metallicity ($\leq 0.2\ \text{Z}_\odot$), and high $L(\text{H}\beta)/L_\nu(1500\text{\AA})$ ratio suggest that the galaxy hosts a young stellar population (age $\lesssim 10$ Myr).
From the conversion of the \hb\ luminosity into SFR \citep{Kennicutt+98}, we derive a star formation rate of $\sim10\ {\rm M}_\odot/{\rm yr}$.
If we assume that the galaxy star formation activity is limited to the clumps \cite[e.g.][]{Genzel+11,ForsterS+11a,Zanella+15,Zanella+19}, we find that clumps are starbursting, having a ${\rm sSFR}\geq 1.25\times 10^{-8}\ {\rm yr}^{-1}$. 

\item As typical of LAEs \cite[e.g.][]{Shibuya+14b,Hoag+19}, the \lya~is extended and offset with respect to the galaxy FUV emission. 
Besides, the \lya\ spectral profile is redshifted ($\Delta v = 403 \pm 4$~km/s) and asymmetric, as expected in case of outflowing gas. 
The \lya\ radiative transfer modelling \cite[e.g.][]{Gronke+15} estimates the expansion velocity of the outflowing material $\sim 211\pm4$~km/s.
This value is in good agreement with the estimate derived by the analysis of the shape of FUV absorption lines.
We obtain a mass loading factor $\eta \sim 1.8 - 2.4$.
These values are consistent with those found in the hydro-dynamical simulations of clumpy galaxies that assume strong/medium SNe feedback \citep[e.g.][]{Bournaud+14,Fensch+20}.

\item We estimate the molecular gas mass of clumps by considering the Schmidt-Kennicutt relation \citep{Sargent+14} and obtain M$_\text{mol} = (7.19^{+9.41}_{-2.58})\times 10^8\ \text{M}_\odot$ (starburst case). Assuming that the detected outflows are the consequence of star formation feedback, the timescale over which the outflows expel the clumps' gas reservoir is $\lesssim 50$ Myr.
We however highlight that our estimate do not take into account the possibility of inflows that could lengthen the clumps' gas expulsion timescale.
\end{itemize}

The results recovered by this study highlight how high-quality multi-wavelength datasets from state-of-the-art instrumentation are essential tools to investigate the properties of clumpy galaxies and understand the nature, and fate of clumps. Despite the fact that current studies are still limited by the spatial resolution achievable with state-of-the-art instrumentation, in the next years both {\it JWST} and ELT are foreseen to profoundly revolutionise clumps studies opening a new window on the rest-frame optical/NIR properties of clumpy galaxies at redshift $z\geq 2$.

\section*{Data availability}
The data underlying this article will be shared on reasonable request to the corresponding author.

\section*{Acknowledgements}
The work presented in this manuscript is based on observations collected at the European Southern Observatory under the ESO programmes 087.B-0875(A), 60.A-9195(A) and 0102.B-0741(A).
It is also based on observations made with the NASA/ESA Hubble Space Telescope, and obtained from the Hubble Legacy Archive, which is a collaboration between the Space Telescope Science In- stitute (STScI/NASA), the Space Telescope European Coordinating Facility (ST-ECF/ESA) and the Canadian Astronomy Data Centre (CADC/NRC/CSA). 
We thank the anonymous referee for the detailed revision of the manuscript and useful comments.
We also thank Francesca Rizzo, Thierry Fusco and Carlos De Breuck for the valuable discussions regarding the work presented in this paper.
M.G. was supported by NASA through the NASA Hubble Fellowship grant HST-HF2-51409 and acknowledges support from \hst\ grants HST-GO-15643.017-A, HST-AR-15039.003-A, and XSEDE grant TG-AST180036.
E.V. acknowledges funding from the INAF for `Interventi aggiuntivi a sostegno della ricerca di main-stream'.
F.V. acknowledges support from the Carlsberg Foundation Research Grant CF18-0388 `Galaxies: Rise and Death'.
C.C.C. acknowledges support from the Ministry of Science and Technology of Taiwan (MOST 109-2112-M-001-016-MY3).




\bibliographystyle{mnras}
\bibliography{biblio} 


\begin{figure}
    \centering
    \includegraphics[width=\columnwidth]{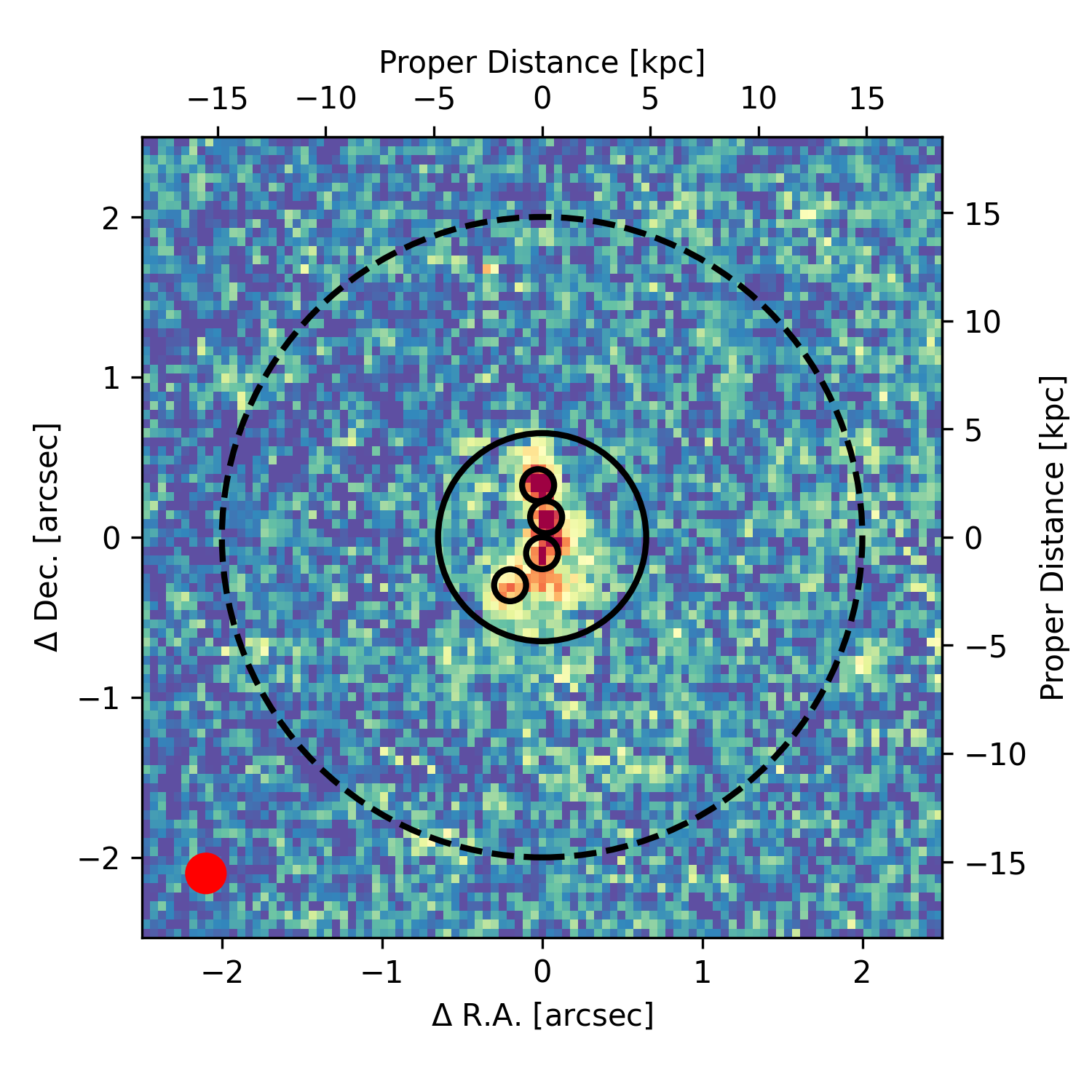}
    \caption{\textit{HST} ACS/WFC F606W cutout of the northern multiple image of our target. The black empty solid circles highlight the position of the 4 detected FUV-bright clumps ($r=0.1''$) and the maximum extent of the galaxy FUV emission ($r=0.65''$), as retrieved from the curve of growth.
    The black empty dashed circle shows the maximum extent within which we reconstructed the curve of growth of the galaxy FUV emission, i.e. $r=2''$. Finally, the red circle in the bottom left corner has a radius equal to the \textit{HST} PSF.}
    \label{fig_appendix:galfit}
\end{figure}

\begin{figure}
    \centering
    \includegraphics[width=\columnwidth]{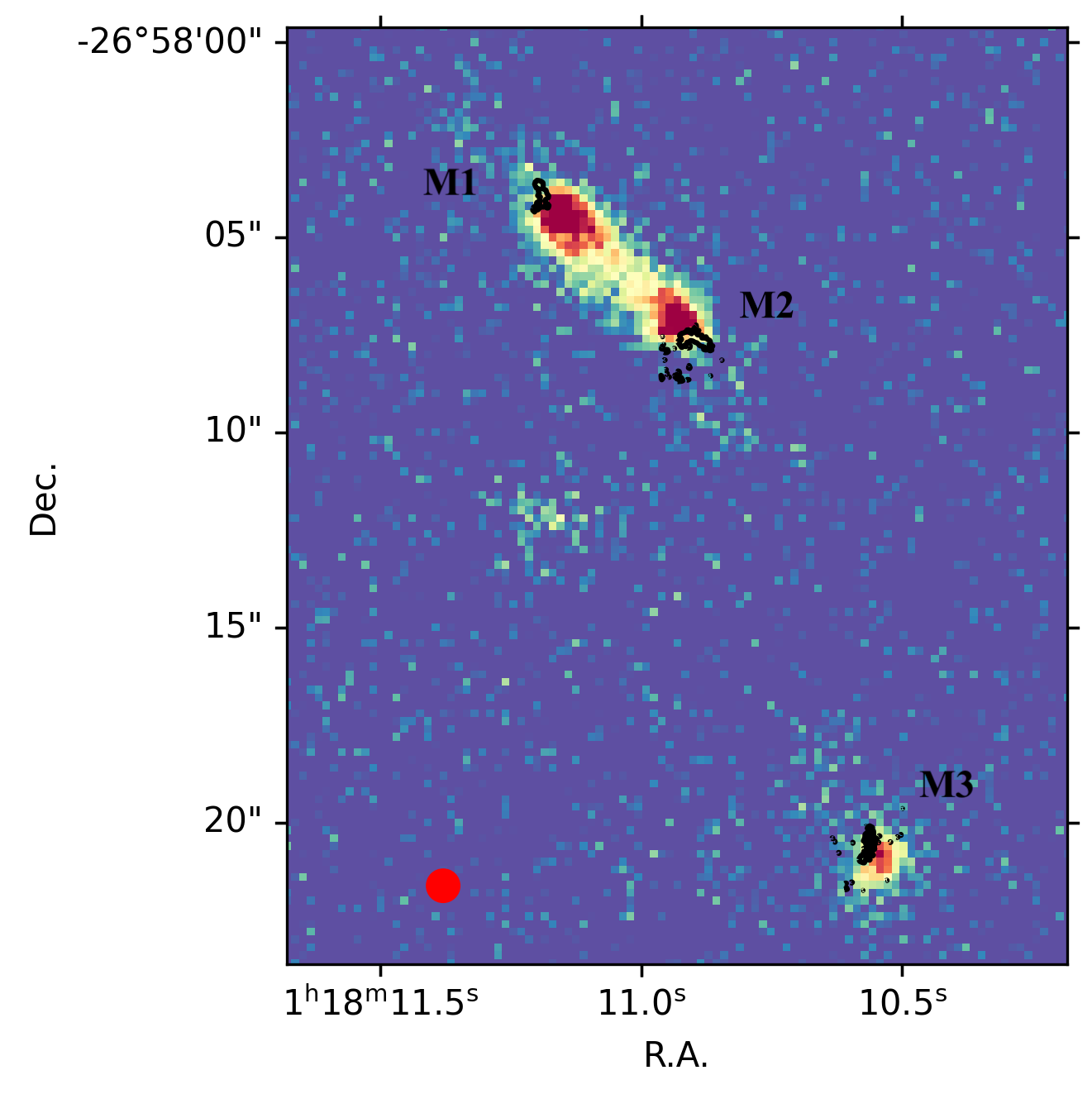}
    \caption{pseudo-NB image of the \lya~emission of our target. The white contours show the position of the three multiple images of the target as detected with {\it HST} in the ACS/WFC F606W filter. The red circle in the bottom left corner has a radius equal to the FWHM of the MUSE PSF ($\sim 0.4''$).}
    \label{fig_appendix:nb_image_lya}
\end{figure}
\begin{figure}
    \centering
    \includegraphics[width=\columnwidth]{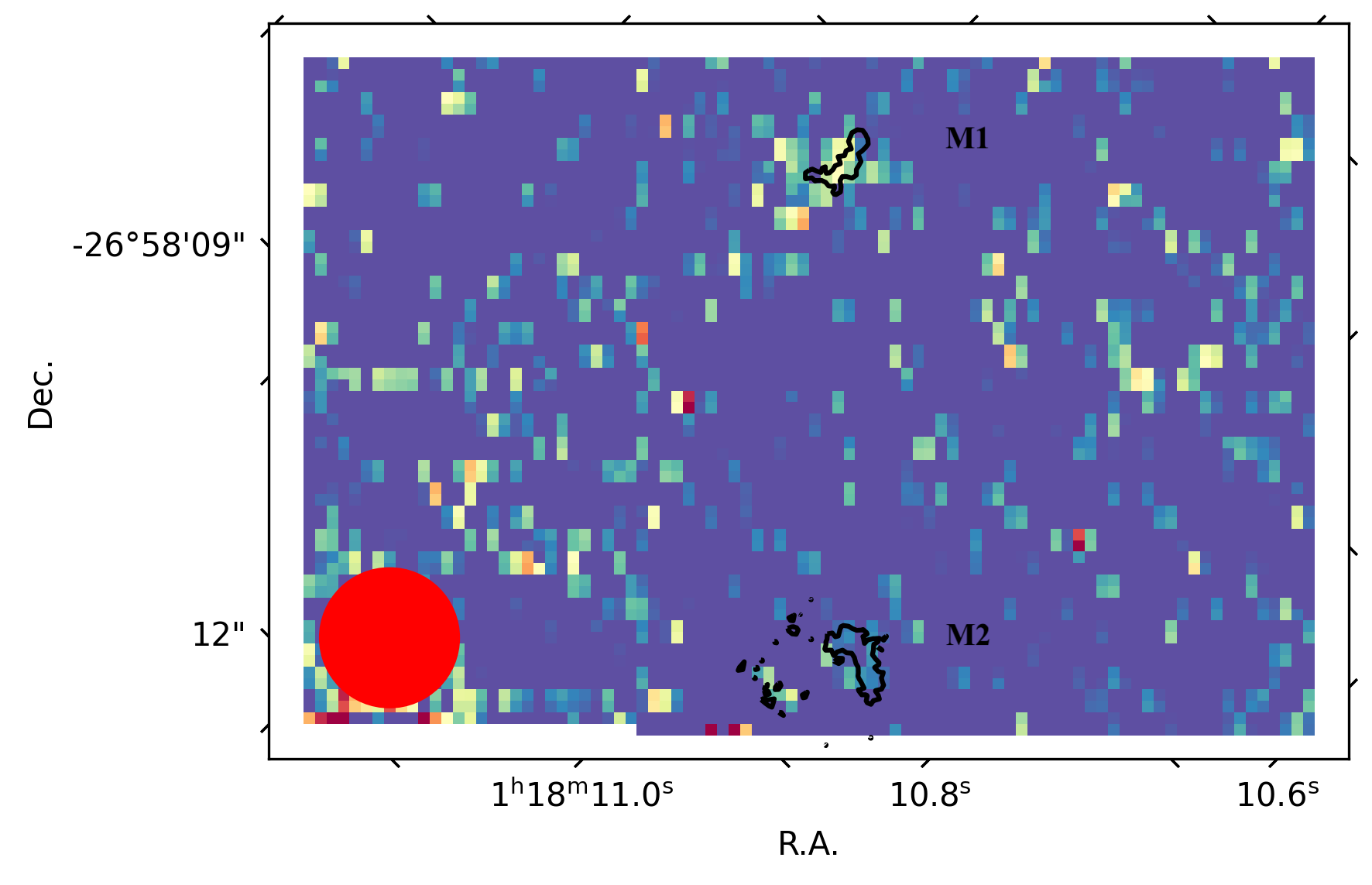}
    \includegraphics[width=\columnwidth]{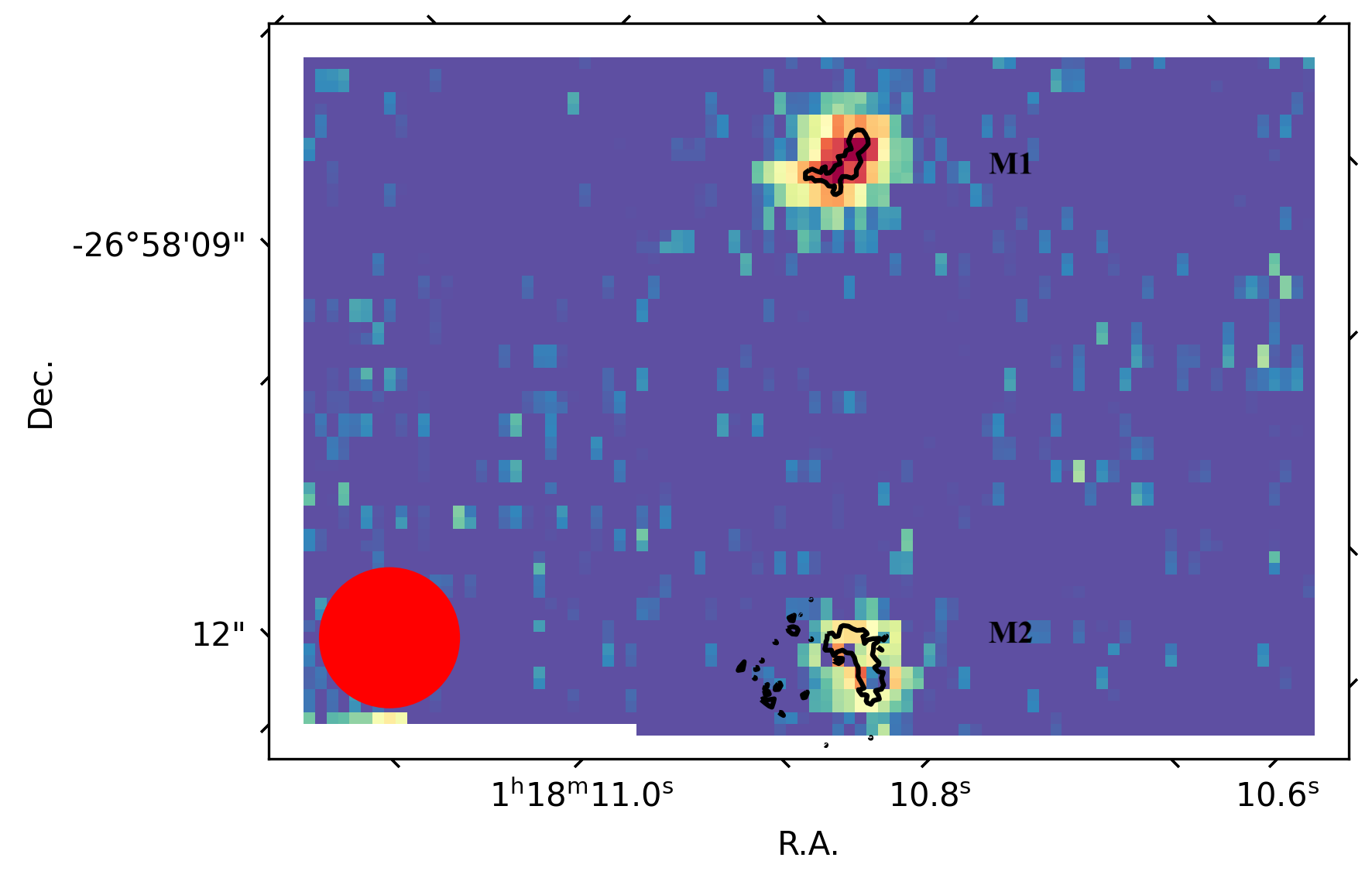}
    \caption{Similar to Figure~\ref{fig_appendix:nb_image_lya} but for the \hb~(top panel) and [OIII]$\lambda 5008$ (bottom panel) emissions. The size of the SINFONI seeing-limited PSF (radius of the bottom left red circle) is of $\sim 0.6''$.}
    \label{fig_appendix:nb_image_sinfoni}
\end{figure}


\begin{figure}
    \centering
    \includegraphics[width=\columnwidth]{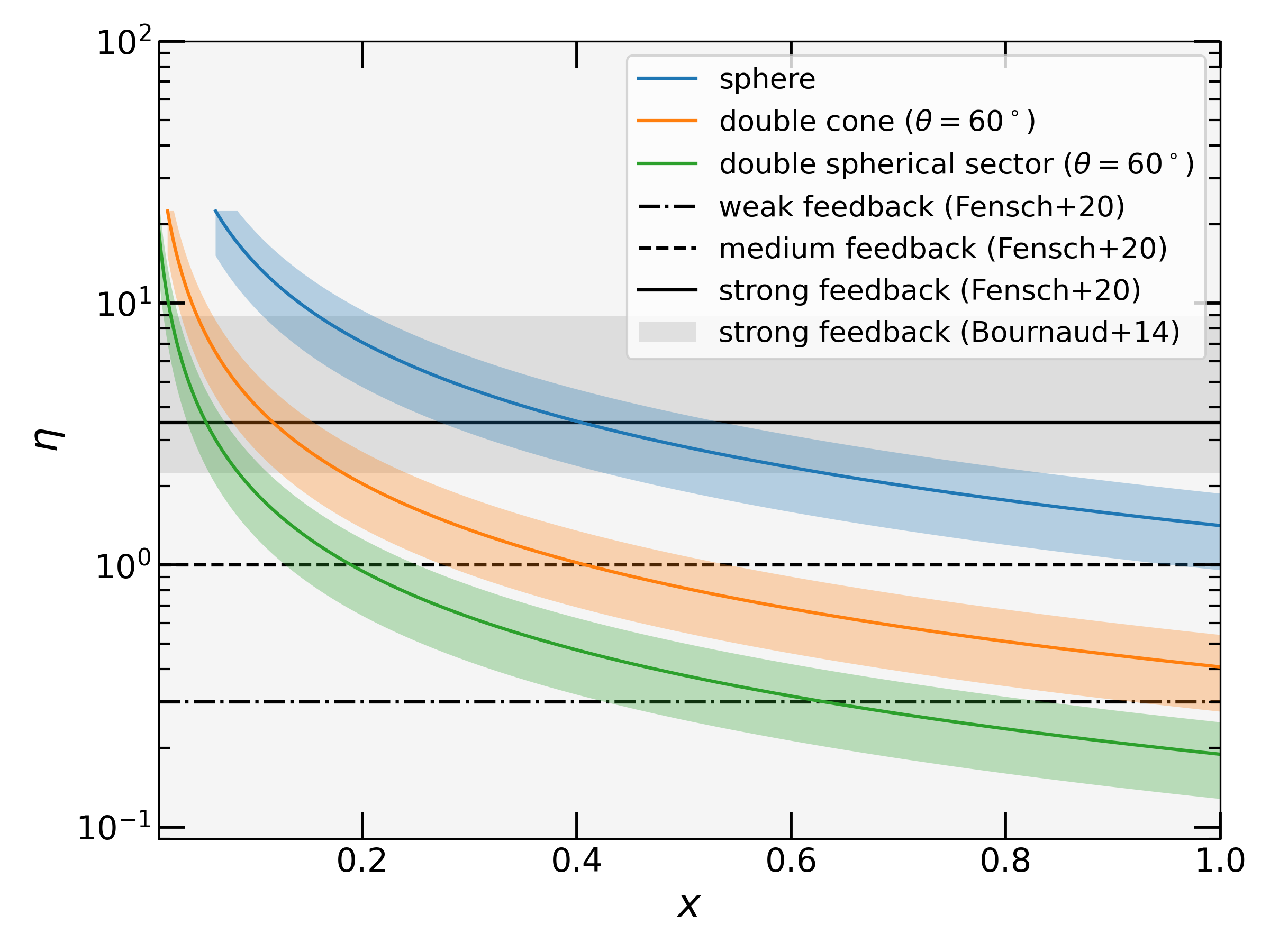}
    \includegraphics[width=\columnwidth]{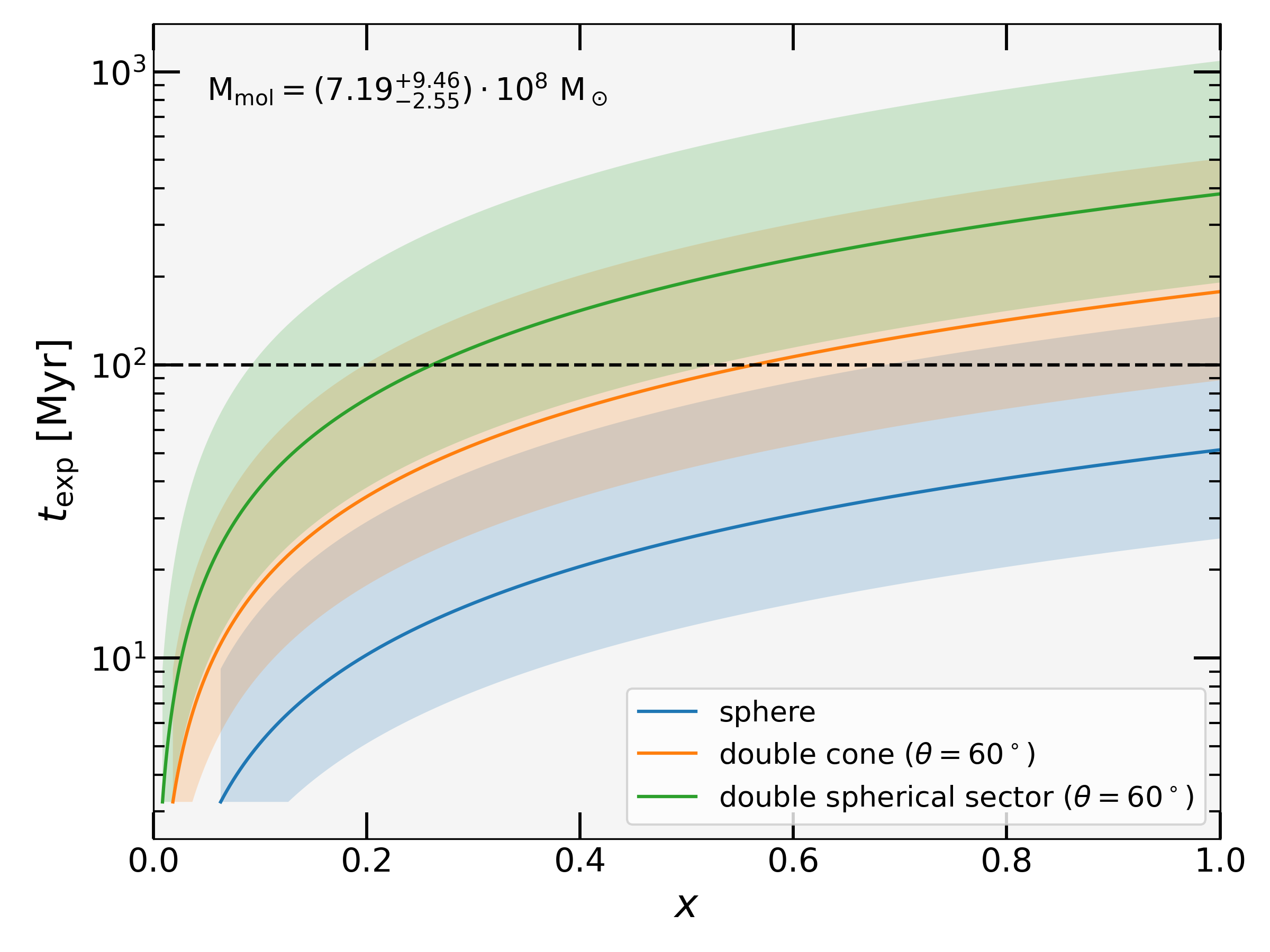}
    \caption{Top panel: Diagram of the mass-loading factor $\eta$ as a function of $x$, and depending on different outflow geometries (i.e. sphere, double cone, double spherical sector). The horizontal black lines show the average values of galaxy-wide $\eta$ that have been found in the simulations by \protect\cite{Fensch+20} implementing weak ($\eta=0.3$), medium ($\eta=1$) and strong ($\eta=3.5$) stellar feedback calibrations, respectively. The horizontal grey shaded area shows the range of $\eta$ that can be obtain only by hydro-dynamical simulations implementing recipes of strong supernovae feedback (e.g. G1, G2 and G3 models) in \protect\cite{Bournaud+14}. Bottom panel: Diagram of the timescale of survival to star formation feedback of clumps ($t_\text{exp}$) as a function of the fraction of HI in the outflow $x$, and depending on different outflow geometries.}
    \label{fig_appendix:clumps_geometries}
\end{figure}

\begin{figure}
    \centering
    \includegraphics[width=\columnwidth]{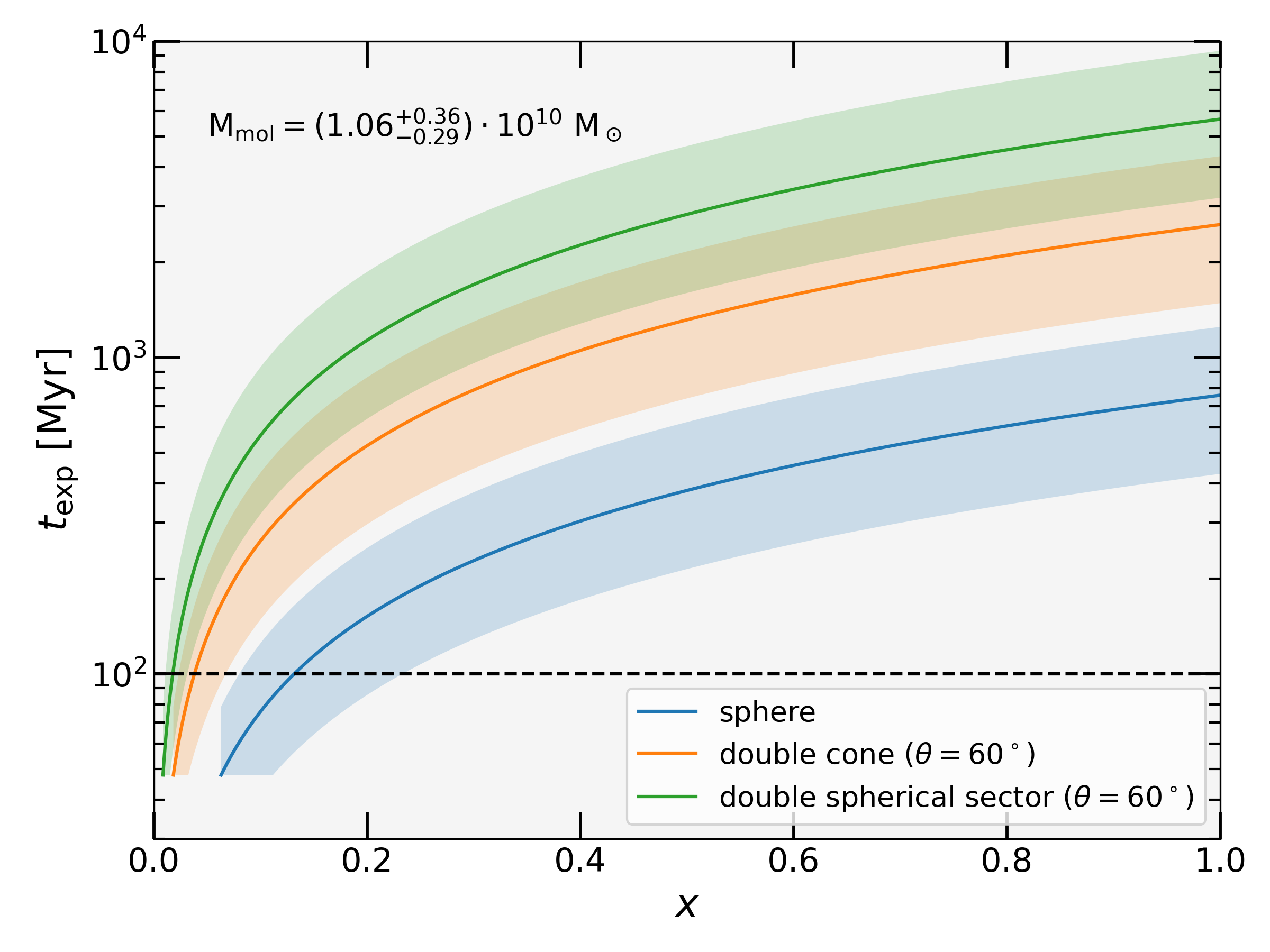}
    \caption{Diagram of the timescale of survival to star formation feedback of clumps ($t_\text{exp}$) as a function of the fraction of HI in the outflow $x$, and depending on different outflow geometries(i.e. sphere, double cone, double spherical sector).}
    \label{fig_appendix:clumps_timescale}
\end{figure}


\appendix
\section{Curve of growth method}
\label{appendix:growth_curve}
In this Section, we report the alternative methodology to the \textsc{galfit} modelling we follow to estimate the contribution of clumps to the total galaxy FUV emission detected by \hst.

As a first step, we estimate the total FUV flux of our target (clumps plus diffuse emission).
To this aim, we consider the BCG-subtracted image and construct a curve of growth measuring the galaxy flux encircled in concentric circular apertures with radii ranging from $0.15''$ to $2''$ (i.e. $\sim 15.8$~kpc).
From the plateau of the curve of growth, we determine the total galaxy flux ($\sim 6.9\times 10^{-19}$  erg/s/cm$^2$/\AA) and size, $r \sim 0.65''$ (i.e. $\sim 4.8$~kpc).

In principle, the total FUV flux of the galaxy measured in the \textit{HST} image could be biased due to the contribution of the \lya~emission that, at the redshift of our target, falls within the ACS/WFC F606W bandpass.
Hence, we compute the contribution of the \lya~emission to the F606W by considering the transmission function of the filter. However, the emission line contribution at the location of the FUV continuum is negligible ($\ll 1$\%).

We estimate the flux of each individual clump by considering non-overlapping apertures with size $r=0.1''$, consistent with the FWHM of the \textit{HST} PSF, see Figure~\ref{fig_appendix:galfit}.
Hence, we apply aperture correction\footnote{
We infer that the energy encircled in a radius of $0.1''$ in the \textit{HST} PSF is of about 67\% with respect to the total. The estimate is in good agreement with what found by \cite{Bohlin+16}, i.e. 66-75\%.
} to the estimated flux taking into account the \textit{HST} PSF.
When summing the flux of all the star-forming regions and comparing it to the total flux of the galaxy, we obtain that clumps constitute 60\% of the \textit{HST} observed light, whereas the remaining 40\% of the UV continuum is likely emitted by a diffuse, low surface brightness component.
The result obtained from the curve of growth method are hence in perfect agreement to those obtained with the \textsc{galfit} modelling, see Section~\ref{subsec:morphology}.


\section{\lya, \hb\ and [OIII]$\lambda5008$ pseudo-NB images}
\label{appendix:pseudo_NB_images}
In this Section, we present the pseudo-NB images of the \lya, \hb, and [OIII]$\lambda5008$ emissions derived following the methodology presented in Section~\ref{subsec:pseudo_nb}.
The pseudo-NB image of the \lya~emission is presented in Figure~\ref{fig_appendix:nb_image_lya}, while the pseudo-NB images of \hb~ and [OIII]$\lambda5008$ are shown in Figure~\ref{fig_appendix:nb_image_sinfoni}.
Because of the wider FoV of MUSE observations ($1'\times1'$), in Figure~\ref{fig_appendix:nb_image_lya} we present a cutout of the MUSE FoV. 
On top of each image, we report the contours of the galaxy FUV emission (in black), as observed with \textit{HST}, and the size of the PSF FWHM (red circle).




\section{Deriving the equation for the mass-loss rate of clumps}
\label{appendix:deriving_mass_loading_factor}
We estimate the gas mass-loss rate ($\dot{M}$) of clumps due to star formation feedback from the equation by  \cite{Pettini+00}:
\begin{equation}
\label{eq:pettini_revisited}
    \dot{M} =  S \cdot n \cdot m_p \cdot v_\text{exp}
\end{equation}
where $S$ is the surface of the expanding region (that depends on the geometry of the outflow), $n$ is the matter density, $m_p$ is the average mass of the particles that constitute the swept up material and $v_\text{exp}$ is the speed of the outflow.  
If we assume that all material within the expanding region is swept up into a shell of thickness $\Delta r_s$ and density $n_s$, we have:
\begin{equation}
    N = n_s \cdot \Delta r_s = \frac{V}{S}\cdot n
\end{equation}
where $N$ is the total column density of the gas within the shell, and $V$ is the volume of the region cleared by the outflow.
Hence, we can rewrite Equation~\ref{eq:pettini_revisited} as:
\begin{equation}
\label{eq:new_mass_loss_rate}
    \dot{M} = \zeta \cdot N \cdot m_p \cdot v_\text{exp}
\end{equation}
where $\zeta = S^2/V$ and depends on the geometry of the outflow.
In our study, we consider three different geometries that could match our observations and are usually adopted when describing feedback solutions: a sphere, a double cone, and a double spherical sector.
Depending on the geometry, $\zeta$ is a function of the distance $r$ swept by the outflowing material and, possibly, the opening angle $\theta$ (only in the biconical and double spherical sector cases). In particular $\zeta$ can be equal to $12\pi r$ (sphere), $6\pi r \tan^2(\theta/2)$ (double cone), or $12\pi r (1-\cos(\theta/2)) $ (double spherical sector).

In Equation~\ref{eq:new_mass_loss_rate}, both $N$ and $m_p$ depend on the chemical composition of the ejected material.
However, we can rewrite the equation as a function of the HI mass ($m_\text{HI}$) and column density ($N_\text{HI}$), introducing a new parameter $x=m_{\rm HI}/m_p$ \cite[e.g.][]{Swinbank+07}. 
Hence, we can write:
\begin{equation}
    \dot{M} =  \frac{\zeta \cdot N_\text{HI} \cdot m_\text{HI} \cdot v_\text{exp}}{x}
\end{equation}
If we express the parameters in the above equation in their typical physical units, we derive the final equation:
\begin{equation}
\label{eq:final_mass_loss}
    \dot{M}~[\text{M}_\odot/\text{yr}] =  \frac{8.19\times 10^{-24}}{x} \cdot \left( \frac{\zeta}{[\text{kpc}]} \right)
    \cdot \left( \frac{N_\text{HI}}{[\text{cm}^{-2}]} \right) \cdot \left( \frac{v_\text{exp}}{[\text{km/s}]} \right)
\end{equation}
In the spherical case (i.e. $\zeta=12\pi r$, the above equation can be written as:
\begin{equation}
\label{eq:final_mass_loss}
    \dot{M}~[\text{M}_\odot/\text{yr}] =  \frac{3.09\times 10^{-22}}{x} \cdot \left( \frac{r}{[\text{kpc}]} \right)
    \cdot \left( \frac{N_\text{HI}}{[\text{cm}^{-2}]} \right) \cdot \left( \frac{v_\text{exp}}{[\text{km/s}]} \right)
\end{equation}

\section{Alternative outflow geometries}
\label{appendix:outflow_geometries}
In this Section, we briefly investigate the impact of the outflow geometry $\zeta$ (see Appendix~\ref{appendix:deriving_mass_loading_factor}) on the estimates of both the mass loading factor $\eta$ and gas removal timescale $t_{\rm exp}$. In particular, we examine the case of bipolar outflows with a biconical and double spherical sector geometry. Similarly to Figure~\ref{fig:outflow_geometries}, we present how both $\eta$ and $t_{\rm exp}$ vary as a function of $x$. As in Section~\ref{subsec:mass_loss}, we set the radius swept by the outflowing medium $r=2.2\pm0.2$~kpc, while we arbitrarily assume an opening angle $\theta=60^\circ$ \cite[e.g.][]{Swinbank+07} since no direct estimates are available based on our dataset. 

For the mass loading factor (top panel of Figure~\ref{fig_appendix:clumps_geometries}), $\eta$ is always grater than unity in the case of a spherical geometry. On the contrary, the biconical and double spherical sector solutions have mass-loss rates comparable to the SFR (or even larger) only for $x\leq 0.55$ and $x\leq 0.20$, respectively, while, in the range of confidence $x=0.6-0.8$, both tracks assume lower $\eta$ values ($\eta=0.2-0.7$). In this case, star formation feedback would be less effective in expelling the gas content of the clumps.

A similar but opposite trend is observed for $t_{\rm exp}$ (bottom panel of Figure~\ref{fig_appendix:clumps_geometries}), since a lowering of $\dot{M}$ translates into an increase of the timescale over which the gas is expelled from the clumps. In this case, while the spherical solution returns $t_{\rm exp}<50$~Myr, the bipolar geometries foresee gas removal timescale up to 400~Myr (100-300~Myr in the $x$-range of confidence).

We highlight how these results are mainly driven by the choice of the outflow opening angle $\theta$. In fact, an increase in $\theta$ brings the tracks closer to the spherical case (the two solutions coincide when $\theta=180^\circ$).

\section{Gas removal timescale in main-sequence clumps}
\label{appendix:clumps_timescale_ms}
In this Section, we report the dependence of the gas removal timescale ($t_\text{exp}$) on $x$, and for three different outflow geometries (spherical, biconical, double spherical sector), in the case of clumps forming stars in a `main-sequence' mode, i.e. supposing that they lie on the stellar mass -- SFR relation of star-forming galaxies \cite[e.g.][]{Elbaz+07,Rodighiero+11,Whitaker+12,Sargent+14} .
In the case of main-sequence clumps, the prescriptions by \cite{Sargent+14} predict a clumps' molecular gas mass M$_\text{mol} = (1.06^{+0.36}_{-0.29})\times 10^{10}\ \text{M}_\odot$, a value $\sim 15$ times higher than the starbursting estimate reported in Section~\ref{subsec:mass_loss}.
Because of the significant increase of M$_\text{mol}$, the tracks of the gas mass removal shift systematically towards longer timescales with the gas being expelled from the clumps by star formation feedback in several hundreds Myr, see Figure~\ref{fig_appendix:clumps_timescale}.
Independently on the geometry and on $x$, main-sequence clumps would retain their molecular gas long enough so that they could contribute to the morphological evolution of the galaxy centre \citep[e.g. bulge growth,][]{Noguchi+99,Genzel+06, Elmegreen+08,Ceverino+10} as the consequence of their migration inward the galaxy disk because of dynamical friction and torques, and coalescence at the centre of the galaxy.

\bigskip
\bigskip
$^{1}$Kapteyn Astronomical Institute, University of Groningen, 9700AV Groningen, The Netherlands\\
$^{2}$Dipartimento di Fisica ed Astronomia, Universit\`a degli Studi di Padova, Vicolo dell’Osservatorio 3, I-35122 Padova, Italy\\
$^{3}$European Southern Observatory, Karl Schwarzschild Straße 2, D-85748 Garching, Germany\\
$^{4}$Istituto Nazionale di Astrofisica (INAF), Vicolo dell’Osservatorio 5, I-35122 Padova, Italy\\
$^{5}$Univ. Lyon, Univ Lyon1, ENS de Lyon, CNRS, Centre de Recherche Astrophysique de Lyon UMR5574, F-69230, Saint-Genis-Laval, France\\
$^{6}$Department of Physics \& Astronomy, Johns Hopkins University, Baltimore, MD 21218, USA\\
$^{7}$School of Mathematics, Statistics and Physics, Newcastle University, Newcastle upon Tyne, NE1 7RU, UK\\
$^{8}$Max-Planck-Institut fur Astrophysik, Karl-Schwarzschild-Str 1, D-85748 
Garching bei M\"unchen, Germany\\
$^{9}$University Observatory Munich (USM), Scheinerstrasse 1, D-81679 Munich, Germany\\
$^{10}$Max-Planck-Institut für extraterrestrische Physik (MPE), Giessenbachstr. 1, D-85748 Garching, Germany\\
$^{11}$Academia Sinica Institute of Astronomy and Astrophysics (ASIAA), No. 1, Section 4, Roosevelt Rd., Taipei 10617, Taiwan\\
$^{12}$Univ. Lyon, ENS de Lyon, Univ. Lyon 1, CNRS, Centre de Recherche Astrophysique de Lyon, UMR5574, 69007 Lyon, France\\
$^{13}$European Southern Observatory,Alonso de Cordova 3107, Vitacura, Santiago, Chile\\
$^{14}$CEA, IRFU, DAp, AIM, Universit\'e Paris-Saclay, Universit\'e Paris Diderot, Sorbonne Paris Cit\'e, CNRS, F-91191 Gif-sur-Yvette, France\\
$^{15}$Centre for Extragalactic Astronomy, Durham University, South Road, Durham, DH1 3LE\\
$^{16}$Institute for Computational Cosmology, Durham University, South Road, Durham DH1 3LE\\
$^{17}$Cosmic Dawn Center (DAWN), Denmark\\
$^{18}$Niels Bohr Institute, University of Copenhagen, Jagtvej 128, DK-2200 Copenhagen N, Denmark\\
$^{19}$Istituto Nazionale di Astrofisica (INAF), Osservatorio di Astrofisica e Scienza dello Spazio, via Gobetti 93/3, 40129 Bologna, Italy\\

\bsp	
\label{lastpage}
\end{document}